\documentclass[aps,prb]{revtex4}
\usepackage{graphicx}
\begin{document}


\title{Fractional quantum Hall effect at $\nu = 5/2$: Ground states,
non-Abelian quasiholes, and edge modes in a microscopic model}

\author{Xin Wan}
\affiliation{Zhejiang Institute of Modern Physics, Zhejiang
University, Hangzhou 310027, P.R. China}

\author{Zi-Xiang Hu}
\affiliation{Zhejiang Institute of Modern Physics, Zhejiang
University, Hangzhou 310027, P.R. China}
\affiliation{National High Magnetic Field Laboratory,
Tallahassee, Florida 32306, USA}

\author{E. H. Rezayi}
\affiliation{Department of Physics, California State University Los
Angeles, Los Angeles, California 90032, USA}

\author{Kun Yang}
\affiliation{National High Magnetic Field Laboratory and Department
of Physics, Florida State University, Tallahassee, Florida 32306,
USA}
\affiliation{Zhejiang Institute of Modern Physics, Zhejiang
University, Hangzhou 310027, P.R. China}

\date{\today}

\begin{abstract}

We present a comprehensive numerical study of a microscopic model of
the fractional quantum Hall system at filling fraction $\nu = 5/2$,
based on the disc geometry.  Our model includes Coulomb interaction and a
semi-realistic confining potential. We also mix in a three-body
interaction in some cases to help elucidate the physics.  We obtain a
phase diagram, discuss the conditions under which the ground state can
be described by the Moore-Read state, and study its competition with
neighboring stripe phases. We also study quasihole excitations and
edge excitations in the Moore-Read--like state. From the evolution of the
edge spectrum, we obtain the velocities of the charge and neutral edge
modes, which turn out to be very different. This separation of
velocities is a source of decoherence for a non-Abelian
quasihole/quasiparticle (with charge $\pm e/4$) when propagating at the
edge; using numbers obtained from a specific set of parameters we
estimate the decoherence length to be around four microns. This sets an
upper bound for the separation of the two point contacts in a double
point-contact interferometer, designed to detect the non-Abelian
nature of such quasiparticles. We also find a state that is a
potential candidate for the recently proposed anti-Pfaffian state. We
find the speculated anti-Pfaffian state is favored in weak confinement
(smooth edge) while the Moore-Read Pfaffian state is favored in strong
confinement (sharp edge).

\end{abstract}

\maketitle

\section{Introduction}

Fractional quantum Hall (FQH) liquids are remarkable many-electron
systems that possesses nontrivial topological order.~\cite{wen95} Such
topological order is reflected in experimental measurable quantities,
including the (fractional) charge and statistics angle of the gapped
quasiparticle/quasihole excitations supported by the system, and the
spectra of gapless edge excitations. By now a large number of
different FQH states, usually labeled by a Landau level (LL) filling
factor $\nu=N_e/N_\phi$ (where $N_e$ is the number of electrons and
$N_\phi$ the number of flux quanta enclosed in the system), have been
observed experimentally. Most of these FQH states are the Laughlin
states or their hierarchy descendents.  These are Abelian FQH states
whose quasiparticles obey Abelian fractional statistics, and their
edge excitations are made of chiral bosonic modes. The quasiparticle charge
has been measured in some of these states,~\cite{fcharge} and
measurement of the statistics angle has been attempted
recently.~\cite{goldman} Edge excitations of such FQH states have also
been probed using electron tunneling~\cite{chang03} and other methods.

Recently much interest and attention have focused on a special FQH
state with filling factor $\nu=5/2$, first observed twenty years
ago.~\cite{willett87} Interest in this system is driven in part by
numerical work~\cite{morf,rh} that suggests the Moore-Read Pfaffian
state~\cite{moore91} is likely realized in the half-filled first
excited Landau level (1LL) at this filling factor.~\cite{note} The
Moore-Read state is qualitatively different from the Laughlin states
and their descendents, in that it is a non-Abelian FQH state, whose
quasiparticles obey non-Abelian statistics~\cite{nayak96} and whose edge
excitations include a branch of {\em fermionic}
mode.~\cite{wen95,milovanovic96} It has been suggested that
non-Abelian quasiparticles can be used for topological quantum
computation,~\cite{kitaev03,freedman02a,freedman02b,dassarma05,bonesteel05}
further fueling the interest in this system.  It is known that the
Pfaffian state is not particle-hole symmetric. It was pointed out very
recently\cite{lee07,levin07} that the particle-hole transformed
Pfaffian state, termed the anti-Pfaffian state, is also a contender at
$\nu=5/2$. These two states are closely related, but different in
important ways that have experimental consequences. While the
electron-electron interaction is particle-hole symmetric when
projected onto a half-filled LL, which suggests these two states would
be degenerate if this were the only term present in the microscopic
Hamiltonian, in reality the degeneracy between the Pfaffian and
anti-Pfaffian states is lifted by terms that break particle-hole
symmetry.  These include LL mixing~\cite{lee07,levin07} and, as we will
show later, confining potential.

In this paper we report results of a numerical study of the half filled
1LL in a disc geometry. Our study is complementary to earlier
numerical works based on the sphere~\cite{morf} and torus~\cite{rh}
geometries, because the disc is the only geometry that allows one to study
edge states and the closely related physics associated with a confining
potential. Our results can be briefly summarized as follows. By
varying both the electron-electron interaction and confining
potential, various types of ground states are stabilized.  We find
that the Moore-Read Pfaffian and possibly the anti-Pffafian ground
states are realized in different regions of the parameter space of our
model.  Within our model they appear to be separated by an
intermediate state that we interpret as a stripe state. We further
study the quasihole and edge excitations of the Pfaffian state, and
show that they indeed have the properties predicted by
theory. Furthermore, we are able to extract the velocities of the
Pfaffian edge modes, which are of importance in addressing both
qualitative and quantitative issues that arise in experimental studies
of the edge states,~\cite{miller07} especially those involving
quasiparticle tunneling in a double point-contact
interferometer.~\cite{fradkin98,stern06,bonderson06a,ardonne07,overbosch07,bishara07}
Some of our results were briefly reported in an earlier
letter.~\cite{wan06}

The rest of the paper is organized in the following way. We describe
our microscopic model and its Hamiltonian (a mixture of Coulomb
interaction and three-body interaction) in Sec.~\ref{sec:model}.  In
Sec.~\ref{sec:groundstate} we study various competing ground states,
which emerge as the lowest energy states in the exact diagonalization
study. In Sec.~\ref{sec:quasihole} we discuss the trapping of charge
$+e/4$ quasiholes by local potentials (generated by, say, an atomic
force microscope tip) in certain ground states which are
supposed to be in the same universality class as the Moore-Read
state. We then discuss the evolution of the edge spectrum with the
variation of
interaction in Sec.~\ref{sec:edgestates}; in particular, we provide an
estimate of the charge and neutral velocities in a real system based
on our model, and discuss the implication in the decoherence in
double point-contact interference experiments.  In
Sec.~\ref{sec:nonabelian} we demonstrate the non-Abelian nature of a
charge $+e/4$ quasihole by comparing the edge spectra of a system with
and without the quasihole.  Potential instability in the fermionic
edge mode is found. In Sec.~\ref{sec:antipfaffian}, we discuss a
potential candidate that emerged from the numerical calculations for the
recently proposed anti-Pfaffian state, and speculate on its stability
conditions. We summarize our results in Sec.~\ref{sec:conclusion}.  We
leave the technical details of the identification of edge states in a
system with mixed three-body and Coulomb interaction to
Appendix~\ref{app:edgemodes}. The detailed analysis of the evolution
of edge states in the pure Coulomb limit is presented in
Appendix~\ref{app:evolution}.

\section{The Microscopic Model}
\label{sec:model}

We consider a microscopic model of a two-dimensional electron gas
(2DEG) confined to a two-dimensional disk, with a mixed Hamiltonian
\begin{equation}
\label{eqn:mixedhamiltonian} H = \lambda H_{3B} + (1 - \lambda)
H_{\rm C}.
\end{equation}
Here, the parameter $\lambda$ interpolates smoothly between the
limiting cases of a purely three-body Hamiltonian $H_{3B}$
($\lambda=1$) and a pure two-body Coulomb Hamiltonian $H_C$
($\lambda=0$). In the following we measure length in units of the
magnetic length $l_B=\sqrt{\hbar c/eB}$ ($B$ is the magnetic field) and
energy in units of $e^2/\epsilon l_B$ ($\epsilon$ is the dielectric
constant), such that all quantities that appear later are in units of
some combination of the two based on their dimensionality.

Explicitly, the three-body interaction $H_{3B}$ has the form
\begin{equation}
\label{eqn:threebody} H_{3B} = -\sum_{i < j <
k}S_{ijk}[\nabla^2_i\nabla^2_j (\nabla^2_i+\nabla^2_j)\delta({\bf r}_i -
{\bf r}_j)
\delta({\bf r}_i -{\bf r}_k)],
\end{equation}
where $S$ is a symmetrizer:
$S_{123}[f_{123}]=f_{123}+f_{231}+f_{312}$, where $f$ is symmetric in its
first two indices.
The $N$-electron Pfaffian
state proposed by Moore and Read~\cite{moore91} for a half-filled
lowest Landau level (0LL)
\begin{eqnarray}
\label{eqn:mooreread}
&&\Psi_{MR} (z_1, z_2, ..., z_N) \nonumber \\
&=& {\rm Pf} \left (1 \over {z_i - z_j} \right ) \prod_{i < j} (z_i -
z_j)^2 \exp \left \{ - \sum_i {|z_i|^2 \over 4} \right \}
\end{eqnarray}
is the exact zero-energy ground state of $H_{3B}$ with the
smallest total angular momentum $M_0 = N(2N-3)/2$.
In Eq.~(\ref{eqn:mooreread}), the Pfaffian is defined by
\begin{equation}
{\rm Pf} M_{ij} = {1 \over 2^{N/2} (N/2)!} \sum_{\sigma \in S_N}
{\rm sgn} \sigma \prod_{k=1}^{N/2} M_{\sigma(2k-1)\sigma(2k)}
\end{equation}
for an $N \times N$ antisymmetric matrix with elements $M_{ij}$.  In
reality, three-body interaction is present due to finite Landau level
mixing. The three-body Hamiltonian also has other zero-energy states,
known as the edge states, which will be discussed in
Sec.~\ref{sec:edgestates}. We note that while the Moore-Read Pfaffian
wave function Eq.~(\ref{eqn:mooreread}) is written for electrons in the
0LL, it is straightforward to generate the corresponding wave
function for electrons in the 1LL by applying the LL raising operator to
every electron. For the rest of the paper we will use the 0LL version of
various wave functions to simplify our discussion, with the
understanding that the 1LL version of the wave function is generated
the same way.

However, there is a more transparent formulation of the three-body
Hamiltonian in terms of projection operators,~\cite{simonetal} which
can be written as
\begin{eqnarray}
H_{3B}=\sum_M \sum_{i<j<k} |\psi_M (i,j,k) \rangle \langle \psi_M (i,j,k)|,
\end{eqnarray}
where $\psi_M (i,j,k)$ is a three-particle wave function specified
below [Eq.~(\ref{eqn:3pwf})] and $M$ is the total angular momentum of
the state. The Hamiltonian for three particles produces a single
non-zero eigenvalue which is unity (provided sufficient number of
orbitals are allowed) as a true projection operator should.  This is
the most natural way to define the scale of the three-body
Hamiltonian.  It is simpler to analyze $\psi_M$ for bosons first.  The
corresponding expression for fermions, as usual, is obtained by
multiplication of an appropriate Jastrow factor.  The Moore-Read wave
function for bosons contains one unit of relative angular momentum in
the Laughlin factor for each pair, instead of two for fermions.  As a
result when three particles are brought together the relative angular
momentum is $2 = 3 \times 1 - 1$ (instead of $5 = 3 \times 2 - 1$).
We now need to project out all relative angular momenta smaller than
two.  In this case, the only possibility is angular momentum zero (see
Ref.~\onlinecite{simonetal} for details). The relative wave function
is thus a constant.  The total angular momentum $M$ will have to be
absorbed by the center of mass wave function, which is
$(z_1+z_2+z_3)^M$. For fermions the normalized wave function is:
\begin{eqnarray}
\label{eqn:3pwf}
\psi_M(z_1,z_2,z_3) = B_M (z_1+z_2+z_3)^{M-3} J(z_1,z_2,z_3),
\end{eqnarray}
where $J(z_1,z_2,z_3)=(z_1-z_2)(z_1-z_3)(z_2-z_3)$,
and the normalization factor is:
\begin{equation}
B_M={1\over (2\pi)^{3/2}}\sqrt{3^{M-4}\over
2^{M+2}(M-3)!}.
\end{equation}
The total angular momentum of the Jastrow factor $J$ is 3 and that
of the center of mass is $M-3$, giving a total angular momentum $M$.

The three-body interaction $H_{3B}$ has a rather simple form in
the occupation space:
\begin{eqnarray}
H_{3B}= && \sum_{m_1>m_2>m_3}\sum_{m_4<m_5<m_6} U(\{m_i\}) \nonumber \\
&\times &
c^\dagger_{m_1}c^\dagger_{m_2}c^\dagger_{m_3}c_{m_4}c_{m_5}c_{m_6},
\end{eqnarray}
and
\begin{equation}
U(\{m_i\})=V(m_1,m_2,m_3)V(m_4,m_5,m_6),
\end{equation}
where $V$ is a completely antisymmetric function of its
arguments. With $M=m_1+m_2+m_3$ we have:
\begin{equation}
V(m_1,m_2,m_3)= \sqrt{(M-1)!\over 2 \times 3^M m_1!m_2!m_3!}{\it A}\{
m_2 m_1 (m_1-1)\},
\end{equation}
and ${\it A}$ is the antisymmetrizer in $m_1,
m_2,\mbox{and}\ m_3$. The difference between the spectra of this form
of the Hamiltonian and the one with the $\delta$-functions is an
overall factor of $\pi^2/8$ in the latter. While there are more
efficient ways to obtain the Moore-Read state~\cite{haldane} that
avoid diagonalizing a three-body Hamitonian, here we need the $H_{3B}$
to generate the spectrum of the mixed three-body and the two-body
Coulomb Hamiltonian.

The Coulomb Hamiltonian $H_C$ includes a two-body Coulomb ($1/r$)
interaction and a one-body confining potential provided by the
neutralizing background charge distributed uniformly on a parallel
disk of radius $R$, placed at a distance $d$ above the 2DEG. This distance
parameterizes the strength of the confining potential, which decreases
with increasing $d$. The rotationally invariant confining potential
comes from the Coulomb attraction between the background charge and
the electrons. Using the symmetric gauge we can write down the
following Hamiltonian for the electrons confined to the 1LL:
\begin{equation}
\label{eqn:chamiltonian} H_{\rm C} = {1\over 2}\sum_{mnl}V_{mn}^l
c_{m+l}^\dagger c_n^\dagger c_{n+l}c_m +\sum_m U_mc_m^\dagger c_m,
\end{equation}
where $c_m^\dagger$ is the electron creation operator for the first
excited Landau level (1LL) single electron state with angular momentum
$m$. $V_{mn}^l$'s are the corresponding matrix elements of Coulomb
interaction for the symmetric gauge, and $U_m$'s the matrix
elements of the confining potential. Additional details of this model
can be found in Ref.~\onlinecite{wan03} where we studied edge
reconstruction of Abelian fractional quantum Hall states at different
$\nu$, including explicit expressions of $U$ and $V$ and an
illustration of the electrostatic configuration associated with $H_C$
(see Fig. 1 of Ref.~\onlinecite{wan03}).

The confining potential we use here is motivated by the
$\delta$-doping technique in 2DEG fabrication. For GaAs/AlGaAs
heterostructures, silicon impurities are deposited in an atomically thick
layer at a distance $d \sim 1000$ \AA\ above the interface where 2DEG
is located to reduce impurity scattering. Therefore, we model the
background potential arising from these ionized silicon impurities,
which ensure charge neutrality in the samples. Even at an
electrostatic level it is clear that $d$ parameterizes the strength of
the confining potential. At small $d$ the potential is strong and also
sharp near the edge, while at large $d$ the potential is weak and
smooth near the edge. Alternatively, one may tune the background
charge density (right at the 2DEG plane) by smearing out the edge
charge density as in an earlier study of edge reconstruction in
integer quantum Hall liquids.~\cite{chamon94} In the study of Abelian
fractional quantum Hall liquids, we find that, e.g., the Laughlin-like
state is stable up to $d \approx 1.5 l_B$, beyond which edge
reconstruction takes place.~\cite{wan03} While we expect the parameter
$d$ appropriately characterizes the confining potential, we note the
detailed sample structures and fabrication processes have an effect on
how realistic the model is.

\begin{figure}
\includegraphics[width=3in]{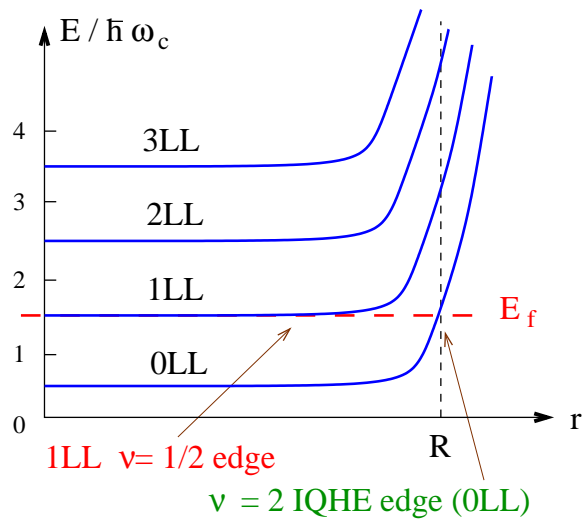}
\caption{\label{fig:landaulevels} (color online).  Illustration
of the edge locations of the $\nu = 1/2$ Moore-Read state
and the $\nu = 2$ integer quantum Hall state. Although the
positive background charge and total electron charge are
5 times that of the electrons forming the Moore-Read state,
the edge is hiding behind the integer quantum Hall edge
in the lowest Landau level. The Moore-Read edge may thus be
protected since it is farther away from the electrostatic edge.  }
\end{figure}

To study the physics at $\nu = 5/2$, we explicitly keep the electronic
states in the half-filled 1LL only, while neglecting the spin up and
down electrons in the lowest Landau level (0LL), assuming they are
inert.  The amount of positive background charge is chosen to be equal
to that of the half-filled 1LL, so the system is neutral. The choice
of a disc radius $R = \sqrt{4N}l_{\rm B}$, where $N$ is the number of
electrons in the 1LL, guarantees that the disk encloses exactly $2N$
magnetic flux quanta, corresponding to $\nu = 1/2$ in the 1LL. This is
a simplification of the real system. In reality, the background charge
equals the {\em total} electron charge of both the half-filled 1LL and
the filled 0LL electrons. The latter neutralizes 4/5 of the background
charge in the bulk, but this neutralization effect is incomplete near
the edge due to finite $d$. Furthermore, the location of the 0LL edge
is different from that of the 1LL electrons (see
Fig.~\ref{fig:landaulevels}).  The physical consequences of these
effects will be discussed in Section ~\ref{sec:conclusion}.  In this
study, we do not consider the finite thickness of the electron layer,
which softens the Coulomb interaction and can be studied using the same
numerical method, albeit time-consumingly.

\section{Competing ground states}
\label{sec:groundstate}

Taking advantage of the rotational invariance of the system, we
diagonalize the Hamiltonian [Eq.~(\ref{eqn:mixedhamiltonian})] for
each Hilbert subspace with a total angular momentum $M$, and 
correspondingly obtain the ground state energy $E(M)$.
The global ground state is defined as the ground state with the lowest
energy $E(M_{gs})$, whose corresponding angular momentum is
$M_{gs}$. In our approach the ground state angular momentum is a
result that comes out of the calculation, rather than a parameter
fixed {\it a priori} based on the property of the state that one is
interested in. Therefore, we can {\em quantitatively} analyze the
stability of the ground state.

\begin{figure}
\includegraphics[width=3in]{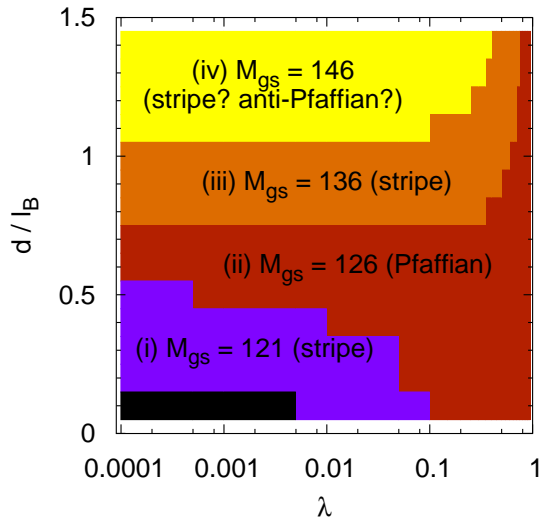}
\caption{\label{fig:phasediagram} (color online).
Total angular momentum of the global ground state as a function of
the mixing parameter $\lambda$ and background charge distance $d$
for 12 electrons in 22 orbitals with the mixed Hamiltonian
[Eq.~(\ref{eqn:mixedhamiltonian})].
The ground state in region (ii) has the same $M_{gs} = 126$
as the Moore-Read (or Pfaffian) wave function. The ground state
in regions (i) and (iii) are believed to be stripe phases.
In region (iv), the ground state is a candidate for the so-called
anti-Pfaffian state (see Sec.~\ref{sec:antipfaffian} for detail).
}
\end{figure}

Figure~\ref{fig:phasediagram} is a phase diagram that shows the total
angular momentum of the global ground state $M_{gs}$ for 12 electrons
in 22 orbitals with the Hamiltonian in Eq.~(\ref{eqn:mixedhamiltonian}).
We vary the mixing parameter $\lambda$ and the background charge
distance $d$. The Moore-Read state for 12 electrons has $M_{\rm MR} =
N(2N-3)/2 = 126$.  In the small $\lambda$ limit, the ground state
around $d = 0.6$-0.7 persist to have $M_{gs} = 126$. To be precise,
the ground state is stable for $0.51 \leq d \leq 0.76$ for the pure
Coulomb case $\lambda=0$.~\cite{wan06} On the other hand, the range
extends as $\lambda$ increases, since three-body interaction favors
the Moore-Read state.  The two regions with $M_{gs} = 121$ and 136
surrounding the Moore-Read ground state are believed to be stripe
phases.
They can be represented by two strings of 0 and 1's $\vert M_{gs} =
121 \rangle = \vert 1000001111111111100000 \rangle$ and $\vert M_{gs}
= 136 \rangle = \vert 1100000001111111111000 \rangle$,
respectively. The 0 and 1's are the occupation numbers of
single-electron angular momentum eigenstates (smaller angular momentum
orbitals to the left). Alternatively, one can understand such a string
as the Slater determinant of the corresponding single-electron angular
momentum eigenstates labeled by 1. At this system size, numerical
ground states have an overlap of about 30-40\% with the corresponding
Slater-determinant states in their range of stability.
For very small
$d$ ($d \approx 0.1 l_B$), $M_{gs}$ can jump to 110 for $\lambda <
0.01$, which is believed to be a finite-size artifact.  On the other
hand, there is a region with ground state $M_{gs} = 146$, which the
authors already showed in Fig.~1b of Ref.~\onlinecite{wan06}.
Toward the pure Coulomb case, this ground state is stable over a range
of $d$ twice as large as that for the Moore-Read state.  We speculate
this is related to the so-called anti-Pfaffian state discussed very
recently.~\cite{lee07,levin07} We will discuss this state in greater
detail in Sec.~\ref{sec:antipfaffian}.

In the absence of three-body interactions, the overlap between the
ground state wave function and the Moore-Read wave function,
$|\langle \Psi_{gs}(M_{gs} = 126)|\Psi_{MR}\rangle|^2$, is about 0.5
for the Coulomb interaction (it jumps up to 0.7 when we tune the $V_1$
pseudopotential~\cite{wan06}). While this is quite substantial
considering the already quite large size of Hilbert subspace, it is
significantly smaller compared with the Laughlin state at $\nu=1/3$ at
comparable system size.  Combined with the narrow window of $d$ within
which $M_{gs} = 126$ in the pure Coulomb interaction case, these
suggest that the Moore-Read state may be quite fragile when system
parameters are varied.  This is consistent with earlier numerical work
on the torus\cite{rh} and the experimental observation that the FQH state
at $\nu = 5/2$ disappears in a tilted magnetic field for modest
tilting angle, even though the state is believed to be spin
polarized. We note the phase boundaries in the small $\lambda$ limit
persist in the small negative-$\lambda$ regime ($-0.02 < \lambda <
0$). This suggests the Moore-Read--like ground state with pure Coulomb
interaction is stable against a small attractive three-body
interaction, which may arise, e.g., due to Landau level mixing.

\section{Non-Abelian quasiholes with electric charge $+e/4$
in the Moore-Read state}
\label{sec:quasihole}

Considering the relatively small overlap and rather narrow window of
stability in $d$, one might wonder if the $M_{gs} = 126$ ground state
is indeed in the same universality class as the Moore-Read state. To
answer this question we must study whether the elementary excitations
of this state have the same properties as those of the Moore-Read
state.  In this section we study the the quasihole excitations
of this state by introducing a local potential, possibly induced in
experiments by the tip of an atomic force microscope, for
example. The next section will be devoted to study of the edge
excitations.

As the ground state of the microscopic model is very sensitive to the
parameters of the system, such as the background confining potential
(by tuning $d$) and the weight ($\lambda$) of the three-body
interaction $H_{3B}$, one may ask if additional features besides the
total angular momentum can offer further support that the ground state
is in the same phase as the Moore-Read state. In fact, one of the most
striking properties of the Moore-Read state is that it supports charge
$\pm e/4$ quasihole/particle excitations, which carry {\em half} the
charge of a Laughlin quasihole/particle at this filling factor.  They
obey non-Abelian statistics, and their existence implies that
electrons are paired in the ground state (in the same way that
observation of $h/2e$ vortices indicate electrons are paired in
superconductors).
We note that the Halperin 331 state~\cite{halperin83} also supports
$e/4$ charge. But it is a bilayer state with 1/4 filling in each
layer, thus $e/4$ charge is not as surprising, as one can get it by
threading a flux quantum through one layer only.

\begin{figure}
\includegraphics[width=3in]{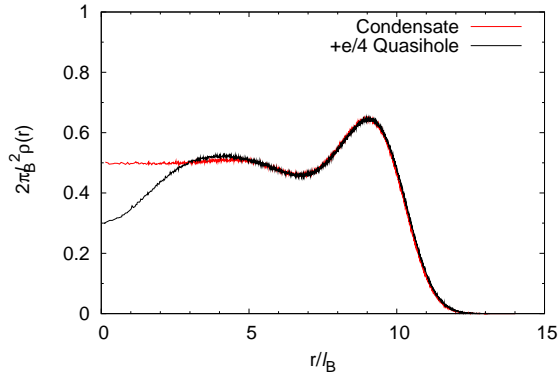}
\caption{\label{fig:qhdensity} (color online).  The densities of the
$+e/4$ Moore-Read quasihole state (Eq.~\ref{eqn:eover4qh}) and
the Moore-Read condensate
(Eq.~\ref{eqn:mooreread}) for a 30-electron system.}
\end{figure}

In an earlier study,~\cite{wan06} we have demonstrated that a
short-range impurity potential at the origin $H_W = W c_0^{\dagger}
c_0$ can induce such a fractionally-charged quasihole, in the presence
of some three-body potential. In a system of 12 electrons in 24
orbitals (as well as a smaller system of 10 electrons in 20 orbitals),
we found for large enough $W$, a quasihole of charge $+e/4$ can appear
at the origin. This is reflected in the depletion of $1/4$ of an
electron in the total occupation number of electrons at orbitals with
small angular momenta, and in the change of ground state angular
momentum from $M_{\rm gs} = N(2N-3)/2$ to $N(2N-3)/2 + N/2$, in
agreement with that of the Moore-Read state with the quasihole located
at the origin:
\begin{eqnarray}
\label{eqn:eover4qh}
&&\Psi_{MR}^{+e/4} (z_1, z_2, ..., z_N) \nonumber \\
&=& {\rm Pf} \left ({z_i + z_j} \over {z_i - z_j} \right )
\prod_{i < j} (z_i - z_j)^2 \exp \left \{ - \sum_i {|z_i|^2 \over 4}
\right \}.
\end{eqnarray}
If $W$ is increased further, a $+e/2$ quasihole (which is a Laughlin
quasihole, equivalent to two $+e/4$ quasiholes~\cite{footnote1})
appears at the origin in the global ground state, whose total angular
momentum further increases to $N(2N-3)/2+N$, in agreement with the
variational wave function
\begin{equation}
\Psi_{MR}^{+e/2} (z_1, z_2, ..., z_N) = \left (\prod_i z_i \right )
\Psi_{MR} (z_1, z_2, ..., z_N).
\end{equation}
In Fig.~\ref{fig:qhdensity}, we show the
electron densities of the $+e/4$ quasihole and the corresponding
Moore-Read ground state for 30 electrons obtained by Monte-Carlo
simulations.  We note that the counterpart of Eq.~\ref{eqn:eover4qh}
on the sphere\cite{footnote} would represent {\em two} $+e/4$
quasiholes on the opposite poles of the sphere.
In the following, we proceed to explore the existence of the $+e/4$
quasihole in a larger parameter space, including cases {\em without}
three-body potential ($\lambda=0$).  Since a $+e/4$ quasihole with
Abelian statistics can arise from a strong-pairing state (instead of
the weak-pairing Moore-Read state), we will discuss the statistics of
the quasihole in Sec.~\ref{sec:nonabelian} after we discuss the edge
excitations of the ground state.

We attempt to trap a quasihole at the origin by introducing a Gaussian
impurity potential:
\begin{equation}
\label{eqn:tippotential}
H_{W,\sigma} = W \sum_m \exp \left ( - m^2 / 2
\sigma^2 \right ) c_m^{\dagger} c_m,
\end{equation}
where $\sigma$ (multiplied by $l_B$) is the range of the
potential. Note $H_W = W c_0^{\dagger} c_0$ is the short-range limit
($\sigma \rightarrow 0$) of the potential in
Eq.~\ref{eqn:tippotential}.  Therefore, the additional parameter
$\sigma$ allows us a more complete search. Two of us and a co-worker
have also been studying the effect of the range and shape of the
potential on the excitation of $\pm e/3$ quasiholes/quasiparticles in
a Laughlin $\nu = 1/3$ liquid.~\cite{hu07}
For $\sigma \sim 2.0$, the weakest strength of the Gaussian potential
that supports the quasihole state as the global ground state is found
insensitive to the confining potential (or $d$ in our
model).~\cite{hu07} In the Moore-Read case, studies also suggest
$\sigma \sim 2.0$ is optimal for the generation of quasiholes, as in
its vicinity the quasihole states can remain to be the global ground
state even in the pure Coulomb case.

\begin{figure}
\includegraphics[width=3in]{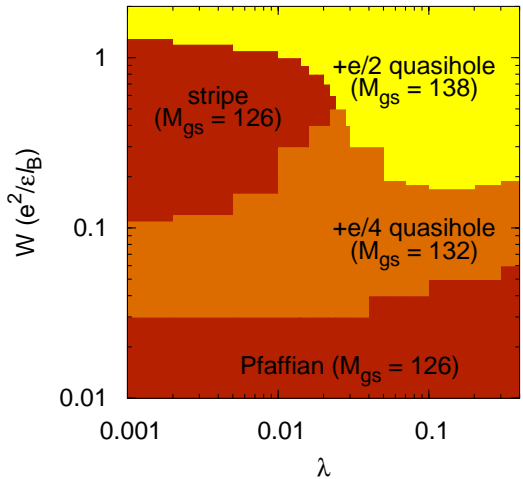}
\caption{\label{fig:quasiholes} (color online).  Ground state angular
momentum as a function of trapping potential and three-body interaction
strength.  We have 12 electrons in 22 orbitals, with the mixed
Hamiltonian [Eq.~(\ref{eqn:mixedhamiltonian})] and the Gaussian tip
potential [Eq.~(\ref{eqn:tippotential})]. $\lambda$ and $W$
characterize the three-body interaction and tip potential strength
respectively.  The background charge is fixed at $d = 0.7 l_B$ above
the electron layer (the ground state has $M_{gs} = 126$, same as the
Moore-Read state in the absence of the Gaussian potential).  For
large enough $\lambda$, as the tip potential strength $W$ increases,
states with $M_{gs} = 132$ or $M_{gs} = 138$, believed to contain a
$+e/4$ quasihole or a $+e/2$ quasihole, become the global ground
state. For small $\lambda$, another ground state with $M_{gs} = 126$
(which is a stripe state with occupation pattern $\vert
0000011111111111100000 \rangle$), separates these two quasihole
states.  }
\end{figure}

Figure~\ref{fig:quasiholes} shows the global ground state angular
momentum as a function of the mixing parameter $\lambda$ and the tip
potential strength $W$ for 12 electrons in 22 orbitals, with the mixed
Hamiltonian in Eq.~(\ref{eqn:mixedhamiltonian}) and the Gaussian tip
potential in Eq.~(\ref{eqn:tippotential}). Here, we choose $d = 0.7
l_B$ and $\sigma = 2.0$. To be specific, we expect the Moore-Read
state with 0, 1, and 2 ($+e/4$) quasiholes to have total angular
momenta of 126, 132, and 138 respectively. For small $W$ ($W < 0.03$),
we find $M_{gs} = 126$ for the global ground state, which is the same
as the Moore-Read state. When there is enough three-body interaction
($\lambda
> 0.025$), as we increase $W$, $M_{gs}$ first jumps to 132 ($W >
0.05$), and then to 138 ($W > 0.2$) as $W$ increases; this is
exactly what one expects when the system first traps a single $+e/4$
quasihole, and then two $+e/4$ quasiholes.  However, for smaller
$\lambda$, there is an additional region with $M_{gs} = 126$ around $W
= 0.5$, separating the one-quasihole and two-quasihole regions.
This region turns out to be a stripe state, characterized by
the binary string $\vert 0000011111111111100000 \rangle$.  Careful
analysis suggests that near $\lambda = 0.025$ and $W = 0.5$, the
energies of the three states with different total angular momenta are
very close to each other and therefore extremely sensitive to the
parameters. Despite this complication, we point out that the trapping of a
single $+e/4$ quasihole by a local potential is a robust feature of
the ground state, which persists to the pure Coulomb case along the
lower boundary ($W \approx 0.03$), at least for finite potential width
of $\sigma = 2.0$. This strongly suggests that the ground state with
$M_{gs} = 126$ is indeed in the universality class of the Moore-Read state.

We note that a $\delta$-function trapping potential
($\sigma\rightarrow 0$ in our Eq. (\ref{eqn:tippotential})) was used
to generate quasiholes on a sphere by T\H oke {\em et al.}~\cite{toke}
They were unable to isolate individual $+e/4$ quasiholes for either the
pure Coulomb or pure three-body interactions, while in our earlier
work\cite{wan06} we succeeded in doing that on a disc for some mixture
of Coulomb and three-body interactions, using the same trapping
potential. One advantage of disc geometry is that one can create a
{\em single} $+e/4$ quasihole in the system, while on a sphere (or
torus) such quasiholes must be created in pairs, and their interaction
complicates the matter. Here we demonstrate that a single $+e/4$
quasihole can also be generated and isolated for pure Coulomb
interaction, with some finite-range trapping potential. We have not,
however, been able to do that with the $\delta$-function trapping
potential. This suggests that such quasiholes have relatively large
size, and its trapping and manipulation will be sensitive to the
details of the trapping potential.  Thus experimentally one may need
to optimize the trapping method in order to generate and manipulate
them.

\section{Edge Excitations of the Moore-Read state and
the Interference Experiments}
\label{sec:edgestates}

In addition to quasihole/quasiparticle properties, another way to probe the
topological order of FQH liquids is to study their edge excitations,
which are also of vital experimental importance. For comparison, the
Laughlin state supports a single branch of bosonic chiral edge mode,
whose properties have been studied in tunneling
experiments.~\cite{chang03} For the Moore-Read state, in addition to a
bosonic mode whose properties are very similar to that of the Laughlin
state, a neutral fermionic branch of excitations has been
predicted;~\cite{wen95,milovanovic96} this fermionic branch is closely
related to the non-Abelian nature of the state.  The existence of both
branches makes the low-energy excitation spectrum of a microscopic
model at $\nu = 5/2$ richer and their experimental consequences more
interesting.~\cite{fendley}

In our earlier study,~\cite{wan06} we have observed both branches of
modes for a mixed Hamiltonian, and demonstrated that a single $+e/4$
quasihole in the bulk changes the boundary condition of the fermionic
mode, clearly indicating the non-Abelian nature of the quasihole. In
this section we provide further details of the analysis of the
spectra, and study how the spectra evolve as the interaction is
varied, especially toward the pure Coulomb interaction.

\subsection{Edge spectrum of a Hamiltonian with mixed
electron-electron interaction}

In this subsection, we demonstrate a clear separation of the fermionic
and bosonic modes for the Moore-Read state,
and try to obtain their velocities for
$\lambda = 0.5$. We will then try to extend the results to the pure
Coulomb case in the next subsection.
We begin by recalling the procedure to
extract edge mode dispersion in the simpler Laughlin case at $\nu = 1/3$,
where there is only one bosonic branch of edge mode. Then, we apply
a similar analysis to the Pfaffian case, where we have a fermionic
branch of edge mode in addition to a bosonic one. Of course, unlike
the Laughlin case, here we need to rely on several reasonable
assumptions, which can be justified {\it post priori}.

In an earlier work,~\cite{wan03} we studied the energy spectrum of
the electron system at $\nu = 1/3$, trying to identify the single
bosonic branch predicted by the chiral Luttinger liquid
theory.~\cite{wen95} The basic idea is that the low-lying excitations
of the quantum Hall system at $\nu = 1/3$ can be described
by a branch of single-boson edge states with angular momentum $l$
($l=1$, 2, 3, ...) and energy $\epsilon_b(l)$. Therefore, we can
label each low-energy state by a set of (bosonic) occupation numbers
$\{n(l)\}$, whose total angular momentum is
\begin{equation}
\label{eqn:angularmomentum}
M = M_0 +
\Delta M = M_0 + \sum_l n(l) l,
\end{equation}
and energy
\begin{equation}
E = E_0 + \Delta E = E_0 +
\sum_l n(l) \epsilon_b(l),
\label{eqn:energy}
\end{equation}
respectively, where $M_0$ and $E_0$ are total angular momentum and
energy of the corresponding ground state.  In Eq. (\ref{eqn:energy}) we
assumed the interactions between the excitations are negligible, which
turns out to be an excellent approximation.  Being edge excitations,
such states
can be independently verified by calculating the squared
matrix elements $T[\{n(l)\}] = | \langle \psi_{\{n(l)\}} (N+1) |
c^{\dagger}_{3N + \Delta M} | \psi_0 (N) \rangle |^2$ numerically in
the microscopic model, and comparing them with the predictions of the
chiral Luttinger liquid theory.~\cite{palacios96,wan03} Note that $M_0
(N+1) - M_0 (N) = 3N$ is the difference in total angular momenta
between the $N$- and ($N+1$)-electron ground states. As shown in
Ref.~\onlinecite{wan03}, even in the presence of background confining
potential, the ansatz of
Eqs. (\ref{eqn:angularmomentum}) and (\ref{eqn:energy}) can be used to
unambiguously identify the bosonic mode energies $\epsilon_b(l)$,
given that edge excitations are not significantly mixed with bulk
excitations. The calculation of $T[\{n(l)\}]$, while not necessary,
does ensure us the correct identification of these excitations as edge
states.

\begin{figure}
\includegraphics[width=3in]{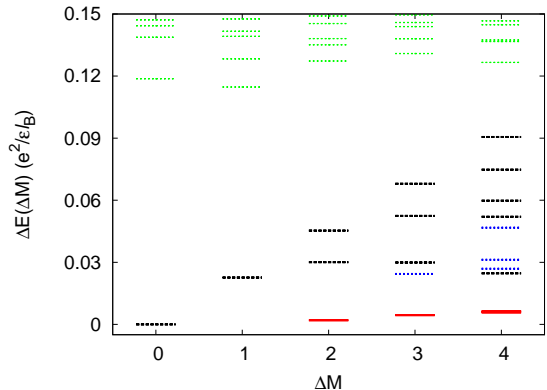}
\caption{\label{fig:edgemodes0.5} (color online).
Low-energy excitations $\Delta E(\Delta
M)$ from exact diagonalization for $N = 12$ electrons
in 26 orbitals in the 1LL (corresponding to $\nu = 1/2$) for the mixed
Hamiltonian in Eq.~(\ref{eqn:mixedhamiltonian}) with $\lambda = 0.5$.
The neutralizing background charge for the Coulomb part is deposited
at $d = 0.6 l_{\rm B}$ above the electron plane.
The red solid bars, the black dashed bars, and the blue dotted bars
mark fermionic, bosonic, and mixed edge excitations, respectively
(see Appendix~\ref{app:edgemodes} for detail). Bulk excitations are
represented by thin dotted bars (green).
}
\end{figure}

Encouraged by the success of identifying the edge mode dispersion and
even predicting the energies of edge excitations in the Laughlin case,
we apply the same analysis to the Moore-Read state. The complication
is that, in addition to the bosonic mode, we also have a fermionic
mode, and thus the convolution of fermionic and bosonic excitations.
Figure~\ref{fig:edgemodes0.5} shows the low-energy excitations $\Delta
E(\Delta M)$ for $N = 12$ electrons in 26 orbitals in the 1LL for the
mixed Hamiltonian with $\lambda = 0.5$ and $d = 0.6 l_B$. A gap at
around $\Delta E = 0.1$ is clearly separating the energy spectrum into
a low-energy section and a higher-energy one.  The numbers of the
low-energy states for $\Delta M = M - M_0 = 0$, 1, 2, 3, and 4 are 1,
1, 3, 5, and 10, respectively, agreeing perfectly with the numbers
expected for the Moore-Read state by earlier theoretical
work.~\cite{wen95,milovanovic96} Notably, the lowest two levels for
$\Delta M = 4$ lie very close to each other. Based on the agreement in
numbers, we are tempted to call them edge states; but further
confirmation comes from the separation and identification of bosonic
mode and fermionic mode, as we show below.

We assume each low-energy excitation can be labeled by two sets of
occupation numbers $\{n_b(l_b)\}$ and $\{n_f(l_f)\}$ for bosonic and
fermionic modes with angular momenta $l_b$, $l_f$, and energies
$\epsilon_b$, $\epsilon_f$, respectively. $n_b(l_b)$ are non-negative
integers while $n_f(l_f)=0, 1$. Since the fermionic edge
excitations are Majorana fermions that obey antiperiodic boundary
conditions,~\cite{milovanovic96} $l_f$ must be positive half
integers, while for bosonic mode $l_b$ are integers. In addition,
the total fermion occupation number $\sum_{l_f} n_f(l_f)$ for
each state must be an even integer because each fermionic excitation
contains an even number of Majorana fermion modes due to their pairing
nature. The angular momentum and energy of the state, measured
relatively from those of the ground state, are
\begin{eqnarray}
\Delta M &=&
\sum_{l_b} n_b(l_b) l_b + \sum_{l_f} n_f(l_f) l_f;\\
\Delta E &=&
\sum_{l_b} n_{\rm b}(l_b) \epsilon_b(l_b) + \sum_{l_f} n_{\rm
f}(l_f) \epsilon_f(l_f).
\end{eqnarray}

\begin{figure}
\includegraphics[width=3in]{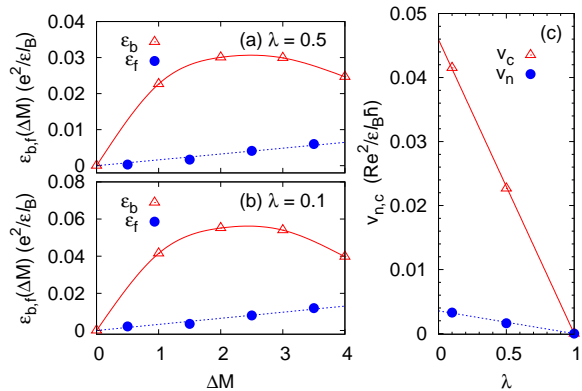}
\caption{\label{fig:dispersion} (color online).  (a) Dispersion curves
of bosonic ($\epsilon_b$) and fermionic modes ($\epsilon_f$) for $N
= 12$ electrons at half filling (in 26 orbitals) in the first Landau
level for $\lambda = 0.5$.  These energies can be used to construct
the complete edge spectrum for the 12-electron system up to $\Delta
M = 4$.~\cite{wan06} (b) Dispersion curves for the same system with
$\lambda = 0.1$, or less three-body interaction. (c) Bosonic ($v_c$)
and fermionic velocities ($v_n$) extrapolated to the pure Coulomb
case. We obtain the $\lambda = 0.5$ and 0.1 points from fitting the
fermionic modes to a straight line in (a) and (b), respectively. At
$\lambda = 1$ (pure three body interaction, $v_n = 0$ as all edge
states have zero energies. We thus obtain $v_c = 0.046$ and $v_n =
0.0036$, in units of $(R e^2)/(\epsilon l_B \hbar)$, for the
pure Coulomb case. }
\end{figure}

The details of the analysis on the data of
Fig. (\ref{fig:edgemodes0.5}) are presented in
Appendix~\ref{app:edgemodes}. Here we summarize the results in
Table~\ref{tbl:edgemodes} and Fig.~\ref{fig:dispersion}(a).
Interestingly, the fermionic dispersion curve is monotonic and can be
well fit by a straight line passing the origin, allowing us to obtain
the neutral fermionic velocity $v_n = d\epsilon_f / dk \approx 0.0016
R e^2/(\epsilon l_B \hbar )$, where the disc radius $R=2\sqrt{N}l_B$,
and we have the conversion from the angular momentum to the linear
momentum along the edge $k = \Delta M / R$.  For typical GaAs systems,
we obtain $v_n \approx 2 \times 10^5$ cm/s.
On the other hand, in contrast to the roughly linear
dispersion of the fermionic branch, the energy of the bosonic branch
bends down (despite a much bigger initial slope or higher velocity),
suggesting a potential vulnerability to edge reconstruction in the
bosonic branch~\cite{wan03,wan02}. This is not surprising since
the bosonic mode is charged; as a result its velocity is dominated
by the long-range nature of the Coulomb interaction in the long-wavelength
limit, but at the same time it is also more sensitive to the
competition between Coulomb interaction and confining potential
which can lead to instability at shorter wavelengths.
If we assume the curve is linear for $k \leq 1 /R$,
we can estimate $v_c \approx 3 \times 10^6$ cm/s for GaAs.

\begin{table}
\begin{center}
\begin{tabular}{cc|cc}
\hline \hline
$l_b$ & \hspace{0.5cm} $\epsilon_b(l_b)$ \hspace{0.5cm}
& \hspace{0.3cm} $l_f$ \hspace{0.2cm} & $\epsilon_f(l_f)$ \\\hline
1 & 0.022659 & \hspace{0.5cm} 1/2 \hspace{0.5cm} &  0.000324 \\
2 & 0.030057 & \hspace{0.5cm} 3/2 \hspace{0.5cm} &  0.001676 \\
3 & 0.029908 & \hspace{0.5cm} 5/2 \hspace{0.5cm} &  0.004117 \\
4 & 0.024668 & \hspace{0.5cm} 7/2 \hspace{0.5cm} &  0.006011 \\
\hline \hline
\end{tabular}
\end{center}
\caption{ \label{tbl:edgemodes} Dispersion energies of both bosonic
and fermionic modes at small momenta for $N = 12$ electrons at half
filling (in 26 orbitals) in the first Landau level. The system has a
Hamiltonian of 50\% Coulomb interaction and 50\% three-body interaction.
The background charge is placed at $d = 0.6$. Based on these
energies, we can construct the complete low-energy (edge) spectrum
of the 12-electron Pfaffian state up to $\Delta M = 4$ (Ref. 26). }
\end{table}

Using these $\epsilon_b$'s and $\epsilon_f$'s (a total of 8 energies),
we can re-construct the whole low-energy spectrum of the system up to
$\Delta M = 4$ (a total of 20 states), which agrees well with the
actual spectrum\cite{wan06} (in fact, we can extend the construction
to $\Delta M = 5$ and obtain very satisfactory agreement for most
states, which do not involve edge modes with larger momentum). The
consistency justifies our analysis based on the assumption that the
interactions between excitations are negligible and further supports
our central result in this section, namely the fermionic mode is well
separated from the bosonic mode and has much lower energy and
velocity.

\subsection{Toward the pure Coulomb interaction}

\begin{figure}
\includegraphics[width=3in]{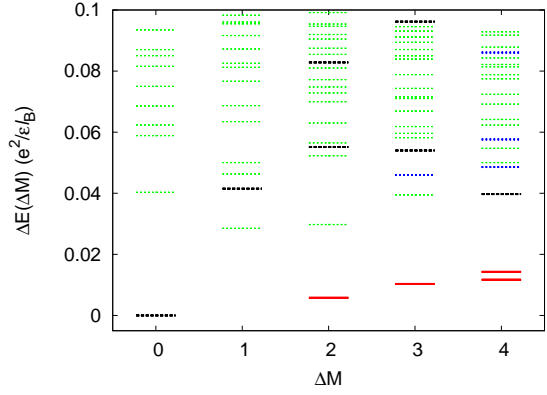}
\caption{\label{fig:edgemodes0.1} (color online).
Low-energy excitations $\Delta E(\Delta
M)$ from exact diagonalization (solid lines) for $N = 12$ electrons
in 26 orbitals in the 1LL (corresponding to $\nu = 1/2$) for the mixed
Hamiltonian in Eq.~(\ref{eqn:mixedhamiltonian}) with $\lambda = 0.1$.
The neutralizing background charge for the Coulomb part is deposited
at $d = 0.6 l_{\rm B}$ above the electron plane.
The red solid bars, the black dashed bars, and the blue dotted bars
mark fermionic, bosonic, and mixed edge excitations, respectively.
Bulk excitations are represented by thin dotted bars (green).
While the bosonic edge excitations mix significantly with the bulk
excitations, the fermionic edge excitations are still well separated
from the rest (see Appendix~\ref{app:evolution0.1} for detail).
}
\end{figure}

\begin{figure}
\includegraphics[width=3in]{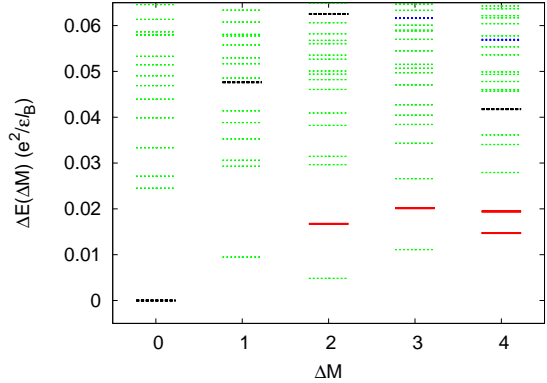}
\caption{\label{fig:edgemodes0.0} (color online).
Low-energy excitations $\Delta E(\Delta
M)$ from exact diagonalization (solid lines) for $N = 12$ electrons
in 26 orbitals in the 1LL (corresponding to $\nu = 1/2$) for
pure Coulomb interaction ($\lambda = 0.0$).
The neutralizing background charge for the Coulomb part is deposited
at $d = 0.6 l_{\rm B}$ above the electron plane.
The red solid bars, the black dashed bars, and the blue dotted bars
mark fermionic, bosonic, and mixed edge excitations, respectively.
Bulk excitations are represented by thin dotted bars (green).
In this case, fermionic edge excitations also mix with the bulk
excitations (see Appendix~\ref{app:evolution0.0} for detail).
}
\end{figure}

The ultimate goal of our work is, of course, to understand the
low-energy spectrum with pure Coulomb interaction, or at least with
less three-body interaction. Looking at the energy spectra for
$\lambda = 0.1$ (Fig.~\ref{fig:edgemodes0.1}) and 0.0
(Fig.~\ref{fig:edgemodes0.0}), we fail to observe a gap separating
edge and bulk states, as in Fig.~\ref{fig:edgemodes0.5}. One
interesting question is, as they start to have similar energies,
whether bulk states and edge states are mixed. However, without the
gap, it is difficult to identify each eigenstate as a bulk state, a
specific edge state, or a mixture of edge and bulk states. To allow
the identification, we calculate the overlaps between the eigenstates
for $\lambda = 0.5$, which we have already analyzed, and the
eigenstates for $\lambda = 0.1$ and 0.0.  We assume the eigenstates
evolve smoothly as $\lambda$ decreases, which turns out to be the case
as our analysis will show. Thus, we can trace the edge states
identified for $\lambda = 0.5$ and sort them out from all eigenstates
in the pure Coulomb case by calculating overlaps; the sorting is
otherwise impossible. In particular, we are interested in the
evolution of fermionic edge states, which play an important role in
understanding the non-Abelian nature of the Moore-Read state.

We leave the details of the approach to Appendix~\ref{app:evolution},
but highlight the main results here. We first look at $\lambda = 0.1$.
Figure~\ref{fig:edgemodes0.1} shows the low-energy excitations for
12 electrons in 26 orbitals in the 1LL for the mixed Hamiltonian
in Eq.~(\ref{eqn:mixedhamiltonian}) with $\lambda = 0.1$.
The neutralizing background charge for the Coulomb part is deposited
at $d = 0.6 l_{\rm B}$ above the electron plane, just as in the previous
case.
We find the fermionic edge excitations (red solid bars) are well
separated from bulk and other edge excitations, as there is clearly a
spectral gap around $\Delta E = 0.02$.
Another observation is for $\Delta M = 4$. Here the two fermionic
excitations
can be significantly mixed with each other, as the two states for
$\lambda = 0.1$ have roughly equal overlap with the two for
$\lambda = 0.5$.

Similar to what we have done in the previous subsection, we can extract
the bosonic and fermionic mode energies for $\lambda = 0.1$, plotted in
Fig.~\ref{fig:dispersion}(b). Interestingly, the figure looks
exactly like that for $\lambda = 0.5$, except the energy scales are roughly
doubled for a higher percentage of Coulomb interaction. The bosonic curve
bends down slightly further. In this case, $v_n (\lambda = 0) = 0.0033
R e^2/(\epsilon l_B \hbar)$.

Figure~\ref{fig:edgemodes0.0} shows the low-energy excitations for pure
Coulomb interaction and the confining potential with $d = 0.6$ for
12 electrons in 26 orbitals. Unfortunately, there is no clear
distinction between edge modes and bulk modes at this system size.
The situation here is similar to a related study on a rotating Bose
gas.~\cite{cazalilla05} After calculating the overlaps, we find the
energies of the lowest fermionic edge excitations are around 0.02, but
there
are lower eigenstates which originate from bulk excitations.
We note that a recent DMRG study
suggest the bulk gap of the fractional quantum Hall liquid at
$\nu = 5/2$ is approximately 0.03.~\cite{feiguin07}

Unfortunately, we can no longer extract meaningful results for the
bosonic and fermionic dispersion curves for the pure Coulomb 
interaction as we did for $\lambda = 0.5$ or
0.1. We believe the reason is that the fermionic mode starts to mix with
bulk states, distorting the dispersion curves (see
Appendix~\ref{app:evolution} for detail).
Nevertheless, we can
extrapolate the bosonic and fermionic velocities from the two
finite-$\lambda$ values, along with the fact that the velocities are
zero for the pure three-body case $\lambda=1$, because all edge states
have zero energy in $H_{3B}$.\cite{milovanovic96} This also suggests
that the velocities should be roughly
proportional to the weight of Coulomb
interaction $1-\lambda$, which is indeed what we find in
Fig.~\ref{fig:dispersion}(c). The extrapolations give $v_c = 0.046$
and $v_n = 0.0036$ for the pure Coulomb case ($\lambda=0$), in units of
$R e^2/(\epsilon l_B \hbar)$.  In GaAs systems they are $v_c \approx
5 \times 10^6$ cm/s and $v_n \approx 4 \times 10^5$ cm/s,
respectively.  We can check the validity of the numbers using
Fig.~\ref{fig:edgemodes0.0}; for example, the bosonic state at $\Delta
M = 1$ is at $E = 0.0476$, very close to 0.046 based on the value of $v_c$.
Similar comparisons also find $v_n \approx 0.004 (R e^2)/(\epsilon l_B
\hbar)$ is in reasonable agreement with the energies at $\Delta M = 2$
and 3.

We close this subsection by noting that the mixing between bulk and
edge excitations seen here for pure Coulomb interaction or small
$\lambda$ is a finite size effect. The edge excitations, which are
gapless in the thermodynamic limit, have a finite gap due to the
existence of a minimum momentum $k$ dictated by system size. When this
gap is larger than the bulk excitation gap (which is quite small for
the Moore-Read--like state with pure Coulomb interaction), mixing
between the
two types of excitations occurs.  They will ultimately separate as
system size increases, in the long-wave length limit $k\rightarrow 0$.
Adding the three-body interaction makes this separation occur at {\em
smaller} system size, by {\em increasing} the bulk gap without
affecting the edge excitation energy much. Thus the effect of adding
three-body interaction is similar to increasing system size, which allows
us to extract useful information within the accessible system sizes.

\subsection{Implications on interference experiments}

Our numerical calculation suggests that the neutral mode velocity
$v_n$ is much smaller than the charge velocity $v_c$. A
similar conclusion has been reached in an effective edge theory study,
which also suggests the neutral velocity has a dynamic
origin.~\cite{yu07} The situation is somewhat similar to what happens
in a 1D Luttinger liquid of spin-1/2 electrons, where the velocity of
the spin mode is in general lower than that of the charge mode, leading to
the so called spin-charge separation. Here we coin a similar term,
``Bose-Fermi separation", to describe the separation of the velocities
of charged bosonic and neutral fermionic edge excitations of the
Moore-Read edge.

In a Luttinger liquid, spin-charge separation is a main source of the
decoherence of a single electron\cite{lehur02}.  Physically this is
because an electron carries both spin and charge; once it enters the
Luttinger liquid however, its spin and charge components propagate
with different velocities, leading to physical separation between the
two after some decoherence time, and loss of integrity of the
electron.

The same physics is relevant to the fate of a charge $\pm e/4$
quasihole/quasiparticle when it is propagating along the Moore-Read
edge. A charge $\pm e/4$ quasihole/quasiparticle carries both a
bosonic component and a ferminic component; the former carries its
charge while the latter is responsible for its non-Abelian
nature. Similar to the case of an electron in a Luttinger liquid, we
expect Bose-Fermi separation to be a main source of decoherence of
such a non-Abelian quasihole/quasiparticle when it propagates at the
edge. This raises a concern that such decoherence may destroy the
interference pattern coming from the interference between charge
$\pm e/4$ quasiholes/quasiparticles in interferometry experiments proposed
recently.\cite{stern06,bonderson06a} In a very recent
work,\cite{bishara07} it was found that the decoherence length is
indeed very sensitive to the velocities:
\begin{equation}
L_{\phi} = { 1 \over 2 \pi k_B T }
\left ( {1/8 \over v_c} + {1/8 \over v_n} \right )^{-1}.
\end{equation}
As a result, in a double point-contact interferometer, the oscillatory
tunneling current due to interference of $\pm e/4$
quasiholes/quasiparticles decays like $I \propto e^{-L/L_{\phi}}$,
where $L$ is the distance between the two point contacts. It is clear
from the equation above that $L_\phi$ is controlled by $v_n$, when
$v_n \ll v_c$, and smaller $v_n$ leads to shorter $L_\phi$.

Based on our numerical results, we can estimate the constraints on the
interferometry experiments due to decoherence. In the pure Coulomb
case, we use the bosonic and fermionic velocities extrapolated in
Fig.~\ref{fig:dispersion}(c).  Assuming the experiments are done at a
temperature of $10$ mK and a magnetic field of $5$ T,~\cite{xia04} we
estimate $L_{\phi} \approx 4$ $\mu$m; this raises concerns on the
appropriate inter--point-contact distance $L$ in interference
experiments.
In fact, this may be a (perhaps overly) optimistic estimate, as we have not
considered the errors due to finite system size and other realistic
issues like filled lowest Landau level. Most importantly, the
confining potential we use in our model (with parameter $d/l_B \sim
1$) is much stronger than that for real systems;\cite{wan02,wan03}
real samples have much bigger $d/l_B$, resulting in weaker confinement
and thus smaller $v_n$, leading to a smaller $L_\phi$ (see next
section for further discussion on this point). Thus our estimate using
parameters extracted from the specific model we use is best viewed as
an upper bound of $L_\phi$. Further investigation on this is thus needed.

We close this subsection by noting that while Bose-Fermi separation
has important consequences on the decoherence of charge $\pm e/4$
quasiholes/quasiparticles, it does not affect charge $\pm e/2$
Laughlin quasiholes/quasiparticles that only carry the Bose
component. The interference pattern due to these Laughlin $\pm e/2$
quasiholes/quasiparticles, unfortunately, does not exhibit the
exciting non-Abelian behavior. Thus in interference experiments it is
possible that while the interference due to charge $\pm e/4$
quasiholes/quasiparticles is lost due to decoherence, one can still
observe an interference pattern due to charge $\pm e/2$
quasiholes/quasiparticles, which is similar to that in Laughlin
states. In addition to Laughlin quasiholes/quasiparticles, there are
also charge $\pm e/2$ quasiholes/quasiparticles that carry a neutral
fermion ($\psi$) but are also Abelian. Bose-Fermi separation does
affect their propagation and thus suppresses their interference. Also
the added fermion component makes tunneling of such $\pm e/2$
quasiholes/quasiparticles irrelevant,\cite{fendley} further reducing
their importance.

\section{Non-Abelian nature of $+e/4$ quasihole and possible
instability of fermionic mode at the edge}
\label{sec:nonabelian}

Now armed with the capability of exciting quasiholes as well as the
knowledge of edge modes, we are in a position to reveal the
non-Abelian nature of a $+e/4$ quasihole by studying the change of
fermionic edge states in the presence of the quasihole.  Such a change
has been reported in an earlier paper by three of the
authors~\cite{wan06} for $\lambda = 0.5$, $d = 0.5 l_B$ and a
short-range tip potential. Here we are presenting a case with less
three-body interaction ($\lambda = 0.1$), weaker confinement ($d = 0.7
l_B$), and a Gaussian tip potential.

\begin{figure}
\includegraphics[width=3in]{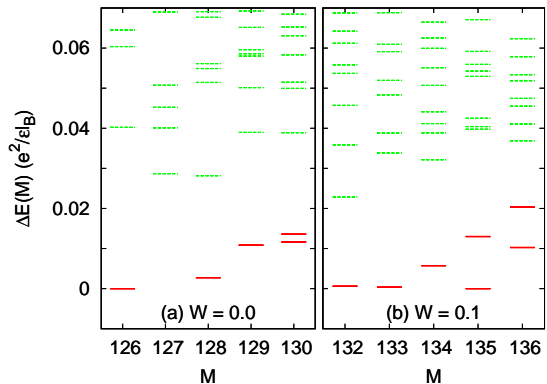}
\caption{\label{fig:nonabelian} (color online).  Low-energy spectra in
a system of 12 electrons in 24 orbitals for the mixed Hamiltonian
with $\lambda = 0.1$ and $d = 0.7 l_B$.  (a) In the absence of the
external tip potential, the ground state ($M = 126$) is
Moore-Read--like.  There are 0, 1, 1, 2 low-energy (fermionic)
excitations for $\Delta M = 1$-4.  (b) In the presence of a Gaussian
tip potential ($W = 0.1$ and $\sigma = 2.0$), the $+e/4$ quasihole
state emerge at $M = 132$.  There are 1, 1, 2, 2 low-energy
(fermionic) excitations for $\Delta M = 1$-4.  This suggests that a
single $+e/4$ quasihole changes the fermionic mode spectrum.  (The
states of interest are marked by red solid bars.) }
\end{figure}

Figure~\ref{fig:nonabelian} shows the low-energy spectra in a system
of 12 electrons in 24 orbitals for the mixed Hamiltonian in
Eq.~(\ref{eqn:mixedhamiltonian}) with $\lambda = 0.1$ and $d = 0.7
l_B$. In the absence of the external tip potential
(Fig.~\ref{fig:nonabelian}a), the ground state ($M = 126$) is
Moore-Read--like, as we can also read from the phase diagram in
Fig.~\ref{fig:phasediagram} (albeit in 22 orbitals). The excitation
spectrum clearly has a gap up to roughly 0.03, consistent with the
result from numerical DMRG calculations.~\cite{feiguin07} Inside the
gap, there are 0, 1, 1, 2 low-energy excitations for $\Delta M = 1$-4
(marked by red solid bars); the numbers agree precisely with the
number of fermionic states, as discussed in
Sec.~\ref{sec:edgestates}. Comparison with Fig.~\ref{fig:edgemodes0.1}
suggests that the fermionic mode dispersion gets distorted by the
increased $d$ (smoother confinement).

In the presence of a Gaussian tip potential ($W = 0.1$ and $\sigma =
2.0$), a new ground state emerges at $M = 132$, reflecting the fact
that a $+e/4$ quasihole has been trapped by the tip potential.
Now there are 1, 1, 2, 2 low-energy (fermionic) excitations for
$\Delta M = 1$-4. For any $M$, there is an energy gap of
at least 0.016 separating the fermionic edge states and the rest. The
results suggest that a single $+e/4$ quasihole changes the fermionic
mode spectrum, a remarkable feature due to the non-Abelian nature of
the $+e/4$ quasihole. More precisely, a quasihole carries a $\sigma$
field of the Ising conformal field theory, and its presence changes
the boundary condition of the edge Majorana fermion mode from being
antiperiodic to periodic. This leads to a shift of the angular
momentum quantum numbers of the lowest energy fermionic edge
excitations.\cite{milovanovic96} Since the non-Abelian properties of
such quasiholes are exclusively due to the $\sigma$ degree of freedom
it carries, observing such a change of boundary condition directly
confirms the non-Abelian nature of the quasihole.

We note that for this particular set of parameters, two of
the low-energy femionic edge states ($\Delta M = 1$ and 3) actually have
very
small {\em negative} energies measured from the single quasihole state
with $M=132$. This is sensitive to the choice of parameters; for
sharper confinement (with a cutoff of 22 orbitals), the quasihole
state {\it is} the global ground state. This does suggest that there
are potential instabilities in the fermionic mode; such possible
instabilities and their consequences remain to be investigated.  We
note that even if such instabilities do not occur, the closeness of
the fermionic excited state energies to zero suggests the neutral
velocity can be even smaller with smoother confinement, which can
jeopardize the estimate we made in the previous section on the dephasing
length relevant to
double point-contact interference experiments.

\section{Discussion on the possible anti-Pfaffian state}
\label{sec:antipfaffian}

In Sec.~\ref{sec:groundstate}, we mentioned a stable ground state
with the same angular momentum quantum number as the recently
proposed anti-Pfaffian state.~\cite{lee07,levin07} In this section,
we discuss how this quantum number is determined.

Suppose we have a system of $N$ electrons in $N_{orb}$ orbitals.  The
Pfaffian state, or the Moore-Read state, has a total angular momentum
of $N(2N-3)/2$. In order to be able to accomodate this state, we need
$N_{orb} \ge 2N-2$. From another angle, we can equivalently view the
system as having $N_h = N_{orb} - N$ holes in $N_{orb}$
orbitals.  The simplest version of the anti-Pfaffian state, by definition,
is the Pfaffian state formed by all the holes; this is possible as
long as $N_{orb} \ge 2N_h-2$, or $N_{orb} \leq 2N + 2$. The total
angular momentum (for holes) is thus
\begin{equation}
M_h = {N_h \over 2}(2N_h-3) = {N_{orb} - N \over 2}[2(N_{orb} - N)-3].
\end{equation}
The total angular momentum in the original electron basis is
\begin{equation}
M_{AP} = N_{orb} (N_{orb} - 1) / 2 - M_h,
\end{equation}
where the first term is the contribution from the electron background
that fully occupies all $N_{orb}$ orbitals, and the hole contribution
$M_h$ is negative because a hole removes an electron from an occupied
orbital.  For $N = 12$ and $N_{orb} = 22$, we find $N_h = 10$, $M_h =
85$, and $M_{AP} = 146$. This is exactly the total angular momentum of
the ground state in the region (iv) in Fig.~\ref{fig:phasediagram}.
Furthermore we found that increasing the three-body interaction enhances
the Pfaffian state and {\em suppresses} this state; this is consistent
with our speculation that this is the anti-Pfaffian state.

This is, however, not definitive evidence, as there are competing
states with the same quantum number. In particular, a 12-electron
stripe-like state represented by the binary string $\vert
1100000000111111111100 \rangle$ has the same angular momentum 146 and
very low
energy.  Analysis of the system with $N = 12$ electrons in 22 orbitals
with pure Coulomb interaction reveals a large overlap (0.35) between
the numerical ground state and the stripe state.

We now explore more general possibilities of the anti-Pfaffian state by
increasing
$N_{orb}$ from 22 to 24.  If the
same 10-hole anti-Pfaffian state were to be realized, the two extra
holes would be at the two
outermost orbitals, and the ground state will have the same quantum
number. Stripe or other states, on the other hand,
are more likely to respond to the change of boundary. In our numerical
calculation, we indeed find the global
ground state still has $M_{tot} = 146$ for $d = 1.2$.  In addition, the
overlap between the ground state and the stripe state discussed above
decreases to
0.13.  This seems to suggest that the stripe phase is favored by
sharp (hard-wall) confinement, as the two outermost orbitals are
unoccupied.  With smoother confinement, a different state, which we
speculate is related to the anti-Pfaffian state, emerges.  At $d =
1.5$, the ground state momentum increases to 151; this can not be
easily explained by a simple stripe phase. On the other hand, it could
be explained as the anti-Pfaffian state with one $+e/4$ quasihole.

These findings suggest that, due to the presence of a confining
potential, the
Pfaffian and anti-Pfaffian states have different energies, and may be
realized
at different confining potential strengths, {\em without invoking
effects of Landau level mixing}.
This is because the confining potential {\em breaks} the particle-hole
symmetry.  Based on
our model calculation, we speculate that the Pfaffian state is stable
as the ground state for strong confinement (sharp edge and smaller
$d$), while the anti-Pfaffian state is stable for weak confinement
(smooth edge and larger $d$). It is worth pointing out that the two
phases are separated and strongly influenced by a stripe
phase. Whether this is a generic feature or a finite-size artifact
cannot be resolved in the current work. Since the anti-Pfaffian is
stable around $d = 1.5 l_B$, it opens another interesting possibility
that edge reconstruction~\cite{wan02} may play a more important role in
the anti-Pfaffian state.

\section{Concluding Remarks}
\label{sec:conclusion}

To summarize, we have studied a microscopic model of fractional
quantum Hall liquids at filling fraction $\nu = 5/2$. The interaction
between electrons are interpolated continuously between the limits of
purely the three-body interaction and purely the Coulomb interaction. Another
parameter we vary in our study is the strength of confinement
potential, parameterized by the distance $d$ separating the positive
neutralizing background charge and the 2D electron gas layer. This
enables us to reveal the nature of ground states and elementary
excitations in the pure Coulomb interaction limit and with
semi-realistic confining potential.  In particular, we find a
Moore-Read--like state is realized in a small window of parameter
space, with predicted properties.

The Moore-Read--like ground state has an edge spectrum consistent with
that of a charged bosonic mode and a neutral fermionic mode. The
fermionic mode has much lower energy than the bosonic mode, implying
the neutral velocity is at least an order of magnitude smaller than
the charge velocity. This leads to a constraint on the dephasing
length for charge $\pm e/4$ quasiholes/quasiparticles: $L_{\phi} < 4$
$\mu$m for typical experimental parameters, at $T=10$ mK. This length
is of crucial importance in double point-contact interference
experiments.

A local potential with a finite width ($\sim 2 l_B$, or about 20 nm),
modeling an atomic force microscope tip, can induce exactly one charge
$+e/4$ quasihole or one charge $+e/2$ (equivalent to 2 charge $+e/4$
quasiholes). From the change of fermionic edge mode when a single
charge $+e/4$ quasihole is excited, we confirm the non-Abelian nature
tof charge $+e/4$ quasihole.

A ground state with the same quantum number as the recently proposed
anti-Pfaffian state is stable in a weak and smooth edge confining
potential. The state is found to be separated from the Moore-Read
Pfaffian state by a stripe-like state in finite-size calculation.

In the present work, we have used a semi-realistic model for the
numerical calculations and attempted to obtain concrete numbers in
experimental units, although further improvement is certainly possible
and probably necessary.  Among the effects we have neglected here,
perhaps the most important is the presence of the electrons occupying
the lowest Landau level (0LL), and their associated edge (see
Fig. \ref{fig:landaulevels}).  These 0LL electrons have two effects
ethat are not included in our study. The first is that the background
charge needs to be equal to the {\em total} electron charge, not just
those in 1LL. While the additional charge is neutralized by the 0LL
electron charge in the bulk, this neutralization is incomplete at the
edge, which results in a fringe electric field\cite{wan03} that tends to
destabilize the 1LL edge through edge reconstruction. On the other
hand, due to the cyclotron gap between the 0LL and 1LL, the 1LL edge
``hides behind" the 0LL edge, and gets protected from instabilities by
the 0LL edge. Thus these two effects that we neglected impact the 1LL
edge in opposite ways, and further studies are needed to resolve which
effect dominates, and the ultimate fate of the 1LL edge.

Nevertheless we do believe the numbers obtained from the present work
can be of use as guidance to experimentalists who are interested in
engineering samples and devices or in manipulating individual
non-Abelian quasiholes in these devices. The parameters considered
here, which describes the smoothness of the edge (related to the
location of $\delta$-doping in realistic epitaxially grown samples)
and the size of an atomic force microscope tip are intimately relevant
to experiments. For example, a momentum-resolved magneto-tunneling
study has found that an epitaxially overgrown cleaved edge can realize
the sharp edge limit.~\cite{huber05} With these realistic issues in
mind, this work supports the possibility of topological quantum
computing~\cite{dassarma07} using fractional quantum Hall states,
although the road ahead needs further exploration.

An immediate follow-up study, which can strengthen the confirmations
found in this work, is the study of the effects of the electron layer
thickness, currently under exploration.  A recent study by Peterson
and Das Sarma~\cite{peterson08} claims that finite layer thickness
enhances the Moore-Read state, using the criterion of wave function
overlap. It would be interesting to study the layer thickness effects
in our more sophisticated model using criteria involving ground state
energy, bulk and edge excitations.  In addition, one also desires to
look at the results in larger systems, where finite-size effects are
weaker. Techniques to reduce the size of the Hilbert space using
various truncation schemes are under development.

\section{Acknowledgments}

We acknowledge the support from NSFC Grant No. 10504028 (X.W.) and NSF
grants No. DMR-0225698, No. DMR-0704133 (K.Y.) and DMR-0606566
(E.H.R.). This research was supported in part by the PCSIRT (Project
No. IRT0754) and by the PKIP of CAS.  Z.X.H. thanks the CCAST for
hospitality during a joint workshop with the KITPC on ``Topological
Quantum Computing'' in Beijing.

\appendix

\section{Analysis of edge excitations at $\lambda = 0.5$}
\label{app:edgemodes}

In this appendix, we discuss in detail the analysis of the edge
excitations for the mixed Hamiltonian with $\lambda = 0.5$, plotted in
Fig.~\ref{fig:edgemodes0.5}. Below $E = 0.1$, we have 1, 1, 3, 5, and
10 states with respective angular momenta $\Delta M= M - M_0= 0, 1, 2, 3, 4$,
which are well separated from the rest; we identify them as low-energy
excitations below the bulk excitation gap. The sequence of numbers are
those expected from edge excitations made of a chiral bosonic branch
and a chiral fermionic branch. Therefore, we want to associate each of
the 20 states with two sets of occupation numbers $\{n_b(l_b)\}$ and
$\{n_f(l_f)\}$ for bosonic and fermionic modes with angular momenta
$l_b$, $l_f$, and energies $\epsilon_b$, $\epsilon_f$, respectively.

Besides the ground state, it is not difficult to identify the only
low-lying state at $\Delta M = 1$ as the bosonic mode with energy
$\epsilon_{\rm b}(1) = \Delta E(\Delta M = 1) = 0.022659$. We can thus
identify all edge states at energies $\Delta E = n_b(1)
\epsilon_b(1)$ with corresponding momenta $\Delta M = n_b(1)$.

For
$\Delta M = 2$, we associate the highest-energy state with $\Delta E
\approx 2 \epsilon_{\rm b}(1)$. There are two more states left, with
energies $\epsilon_b(2)$ and $\epsilon_f(1/2) + \epsilon_f(3/2)$.
There are thus two choices. However, given $\epsilon_b(1) \approx 0.02$,
it is reasonable to assume $\epsilon_b(2) = 0.030057$ is the higher
one of the two. As a result, the fermionic state with the smallest
momentum has much lower energy than the bosonic ones. Counting the
energy states with nearly zero energy (or to be more precise, with
$\Delta E < 0.01$), we find 0, 1, 1, and 2 states for $\Delta M =
1$, 2, 3, and 4, respectively. These numbers agree perfectly with
the results expected for a single branch of Majorana fermion
mode.~\cite{wen95,milovanovic96} We thus assume these energies are
sums of two Majorana fermion energies. For $\Delta M = 2$, for example,
we have already assumed $\Delta E = \epsilon_f(1/2) + \epsilon_{\rm
f}(3/2)$ for the only state.

For $\Delta M = 3$, we have five
states. We continue to assume that the lowest one is a purely fermionic state
with $\Delta E = \epsilon_f(1/2) + \epsilon_f(5/2)$. We can also
identify two bosonic states with energies $\Delta E = 3 \epsilon_b
(1) \approx 0.066$ and $\epsilon_b (1) + \epsilon_b (2) \approx
0.052$. We also find one more from the convolution of both bosonic
and fermionic modes with energy $\Delta E = \epsilon_b (1) +
\epsilon_f(1/2) + \epsilon_f(3/2) \approx 0.024$. The edge state
left should then be the bosonic state with $\Delta E = \epsilon_b
(3) = 0.029908$.

The situation becomes more complicated at $\Delta M = 4$, where we
have two fermionic, five bosonic, and three convoluted edge
excitations. It is easy to identify the convoluted excitations
first, at energies $\epsilon_b(1) + \epsilon_f(1/2) +
\epsilon_f(5/2)$, $\epsilon_b(2) + \epsilon_{\rm f}(1/2) +
\epsilon_f(3/2)$, and $2 \epsilon_b(1) + \epsilon_f(1/2) +
\epsilon_f(3/2)$. Using $\epsilon_b(l)$ for $l = 1$-3 obtained
above, we can identify four bosonic states at energies $4\epsilon_
b(1)$, $2\epsilon_b(1) + \epsilon_{\bf b}(2)$, $2\epsilon_b(2)$, and
$\epsilon_b(1) + \epsilon_b(3)$. The only state with energy $\Delta E
> 0.01$ is thus the remaining bosonic state with $\epsilon_b(4) =
0.024668$. Once again, the two fermionic states have much smaller
energy $\Delta E = \epsilon_f(1/2) + \epsilon_f(7/2)$ and
$\epsilon_f(3/2) + \epsilon_f(5/2)$. We note that in order to write
down a variational wave function for a pair of Majorana-Weyl
fermions with momenta $l > k$, we need at least $(2N+l-1)$ orbitals.
Therefore, by reducing the Hilbert space by using fewer orbitals, the
hard-wall edge confinement will increase some fermionic mode energies,
but leave others intact. This is a test that can unambiguously
distinguish the two states. In particular, $\epsilon_f(7/2)$
will suffer from an energy increase when we reduce the total
number of orbitals to 25, while $\epsilon_f(5/2)$ will remain roughly
unchanged unless we further reduce the orbital number to 24 and
below. We have observed this confinement effect in numerical
calculations, which suggests the state with energy $\epsilon_f(3/2) +
\epsilon_f(5/2)$ is the lower of the two. This energy, together with
the two fermionic excitations at smaller momenta, allow us to solve
$\epsilon_f(1/2)$, $\epsilon_f(3/2)$, and $\epsilon_f(5/2)$.
Consequently, the energy of the other state [$\epsilon_f(1/2) +
\epsilon_{\rm f}(7/2)$] allows us to solve for $\epsilon_f(7/2)$. The
results are summarized in Table~\ref{tbl:edgemodes} and
Fig.~\ref{fig:edgemodes0.5}.

Interestingly, the fermionic dispersion curve is smooth, monotonic and
can be well fit by a straight line passing the origin, allowing us to
obtain the neutral fermionic velocity $v_n$.  The bosonic dispersion
curve, on the other hand, is non-monotonic and bends downward, which
indicates a tendency toward edge reconstruction.~\cite{wan02} A very
similar analysis can be perfromed for $\lambda = 0.1$, which we leave
out for brevity. The results are compared with $\lambda = 0.5$ in
Fig.~\ref{fig:dispersion}.

\section{Numerical analysis of the evolution of edge excitations}
\label{app:evolution}

\subsection{$\lambda = 0.1$}
\label{app:evolution0.1}

\begin{table}
\begin{center}
\begin{tabular}{lccccc}
\hline\hline
\hspace{0.5cm}$\Delta M = 1$\hspace{0.5cm}
& \multicolumn{5}{l}{ $\lambda = 0.1 \ \longrightarrow$ } \\ 
$\downarrow \ \lambda = 0.5$
& \hspace{0.7cm}1\hspace{0.7cm} 
& \hspace{0.7cm}2\hspace{0.7cm}
& \hspace{0.7cm}3\hspace{0.7cm}
& \hspace{0.7cm}4\hspace{0.7cm} 
& \hspace{0.7cm}5\hspace{0.7cm}
\\ \hline
\hspace{0.7cm}1 & 0.017 & \underline{0.827} & 0.034 & 0.001 & 0.000 \\
\hspace{0.7cm}2 & 0.594 & 0.048 & 0.266 & 0.000 & 0.004 \\
\hspace{0.7cm}3 & 0.230 & 0.003 & 0.470 & 0.112 & 0.068 \\
\hspace{0.7cm}4 & 0.004 & 0.000 & 0.033 & 0.361 & 0.278 \\
\hspace{0.7cm}5 & 0.000 & 0.056 & 0.002 & 0.003 & 0.026 \\
\hline\hline
\end{tabular}
\end{center}
\caption{ \label{table:N12.Norb26.Mtot127.D0.6.M3B0.5vs0.1.1LL}
Overlap matrix of the two systems with $\lambda = 0.5$ (row) and 0.1
(column) for $N = 12$, $N_{orb} = 26$, $M = 127$, and $d = 0.6$. The
largest overlap between eigenstates for $\lambda = 0.1$ and the
lowest state (edge state) for $\lambda = 0.5$ comes from the second
lowest state for $\lambda = 0.1$, with a value of 0.827
(underlined). }
\end{table}

We discuss here in detail how we identify the edge excitations
in Fig.~\ref{fig:edgemodes0.1} by calculating the overlaps
between eigenstates for different $\lambda$.
Let us start with the simplest nontrivial case $\Delta M = 1$
(Table~\ref{table:N12.Norb26.Mtot127.D0.6.M3B0.5vs0.1.1LL}). The
lowest excitation (state \#1) for $\lambda = 0.5$ has the largest
overlap (0.827, underlined) with the second excitation (state \#2)
for $\lambda = 0.1$. Meanwhile, state \#1 for $\lambda = 0.1$ has
large overlaps with states \#2 and \#3 for $\lambda = 0.5$, both
bulk excited states. Therefore, we can identify
$(\Delta M, \Delta E) = (1, 0.0415)$ (state \#2 for $\Delta M = 1$)
as an edge state for $\lambda = 0.1$,
with an overlap of 0.827 with the corresponding
edge state \#1 for $\lambda = 0.5$.

For $\Delta M = 2$, we find that states \#1, \#4, and \#12 have
significant overlaps with the lowest three edge states for for
$\lambda = 0.5$, as listed in
Table~\ref{table:N12.Norb26.Mtot128.D0.6.M3B0.5vs0.1.1LL}. As in
Fig.~\ref{fig:edgemodes0.1}, $\Delta E (\#12, \Delta M = 2) = 0.0828$
is roughly the sum of two $\Delta E(\#2, \Delta M = 1) = 0.0415$. This
simple addition law resembles the one we have found for edge states at
one of the Laughlin filling fractions $\nu = 1/3$,~\cite{wan03}
reflecting the conservation of energy and angular momentum.

We may
assume, based on this observation, that $\Delta E (\Delta M = 3) = 3
\Delta E(\#2, \Delta M = 1)$ is an edge state (not plotted in
Fig.~\ref{fig:edgemodes0.1}). The other four edge
states for $\Delta M = 3$ are found, by comparing overlaps, to be
states \#1, \#3, \#4, and \#20, according to
Table~\ref{table:N12.Norb26.Mtot129.D0.6.M3B0.5vs0.1.1LL}.
We note the approximate equalities $\Delta E (\#20, \Delta M = 3) \approx
\Delta E(\#2, \Delta M = 1) + \Delta E(\#4, \Delta M = 2)$ and
$\Delta E (\#3, \Delta M = 3) \approx \Delta E(\#2, \Delta M = 1) +
\Delta E(\#1, \Delta M = 2)$.

Similarly, we can identify six edge excited states in
the lowest 20 eigenstates we have calculated for $\Delta M = 4$. In
addition, we can postulate the existence of another four edge states
with excitation energies of $4 \Delta E(\#2, \Delta M = 1)$, $2
\Delta E(\#2, \Delta M = 1) + \Delta E(\#4, \Delta M = 2)$, $2
\Delta E(\#4, \Delta M = 2)$, and $\Delta E(\#2, \Delta M = 1) +
\Delta E(\#4, \Delta M = 4)$, respectively. Again, the simple
conservation law seems to work fairly well.
We point out that the two fermionic edge states (\#1 and \#2),
whose energies are close, mix significantly with each other
with respect to the $\lambda = 0.5$ case.
To a lesser extent, another two states (\#3 and \#4) also
mix with each other.

\begin{table}
\begin{center}
\begin{tabular}{lcccccc}
\hline\hline
\hspace{0.5cm}$\Delta M = 2$\hspace{0.5cm}
& \multicolumn{5}{l}{ $\lambda = 0.1 \ \longrightarrow$ } \\ 
$\downarrow \ \lambda = 0.5$
& \hspace{0.5cm}1\hspace{0.5cm} 
& \hspace{0.5cm}2\hspace{0.5cm}
& \hspace{0.5cm}3\hspace{0.5cm}
& \hspace{0.5cm}4\hspace{0.5cm} 
& \hspace{0.4cm}$\cdots$\hspace{0.4cm}
& \hspace{0.5cm}12\hspace{0.5cm} 
\\ \hline
\hspace{0.7cm}1 & \underline{0.910} & 0.013 & 0.001 & 0.000
& $\cdots$ & 0.000 \\
\hspace{0.7cm}2 & 0.000 & 0.011 & 0.169 & \underline{0.638}
& $\cdots$ & 0.000 \\
\hspace{0.7cm}3 & 0.000 & 0.000 & 0.000 & 0.000 & $\cdots$
& \underline{0.730} \\
\hspace{0.7cm}4 & 0.002 & 0.753 & 0.001 & 0.033 & $\cdots$ & 0.000 \\
\hspace{0.7cm}5 & 0.003 & 0.021 & 0.464 & 0.104 & $\cdots$ &
0.013 \\
\hline\hline
\end{tabular}
\end{center}
\caption{ \label{table:N12.Norb26.Mtot128.D0.6.M3B0.5vs0.1.1LL}
Overlap matrix of $\lambda = 0.5$ (row) and 0.1 (column) for $N =
12$, $N_{orb} = 26$, $M = 128$, and $d = 0.6$. The underlined
numbers are the overlap between an edge state for $\lambda = 0.5$ and the
(likely) corresponding edge state for $\lambda = 0.1$.}
\end{table}

\begin{table}
\begin{center}
\begin{tabular}{lcccccc}
\hline\hline
\hspace{0.5cm}$\Delta M = 3$\hspace{0.5cm}
& \multicolumn{5}{l}{ $\lambda = 0.1 \ \longrightarrow$ } \\ 
$\downarrow \ \lambda = 0.5$
& \hspace{0.5cm}1\hspace{0.5cm} 
& \hspace{0.5cm}2\hspace{0.5cm}
& \hspace{0.5cm}3\hspace{0.5cm}
& \hspace{0.5cm}4\hspace{0.5cm} 
& \hspace{0.4cm}$\cdots$\hspace{0.4cm}
& \hspace{0.5cm}20\hspace{0.5cm} 
\\ \hline
\hspace{0.7cm}1 & \underline{0.910} & 0.001 & 0.000 & 0.000
& $\cdots$ & 0.000 \\
\hspace{0.7cm}2 & 0.000 & 0.000 & \underline{0.895} & 0.005
& $\cdots$ & 0.001 \\
\hspace{0.7cm}3 & 0.000 & 0.185 & 0.004 & \underline{0.524}
& $\cdots$ & 0.001 \\
\hspace{0.7cm}4 & 0.000 & 0.000 & 0.000 & 0.000 & $\cdots$
& \underline{0.351} \\
\hspace{0.7cm}5 & 0.000 & 0.000 & 0.000 & 0.000 & $\cdots$
& 0.000 \\
\hline \hline
\end{tabular}
\end{center}
\caption{ \label{table:N12.Norb26.Mtot129.D0.6.M3B0.5vs0.1.1LL}
Overlap matrix of $\lambda = 0.5$ (row) and 0.1 (column) for $N =
12$, $N_{orb} = 26$, $M = 129$, and $d = 0.6$. The underlined
numbers are the overlap between an edge state for $\lambda = 0.5$ and the
(likely) corresponding edge state for $\lambda = 0.1$.}
\end{table}

\begin{table}
\begin{center}
\begin{tabular}{lccccccc}
\hline\hline
\hspace{0.5cm}$\Delta M = 4$\hspace{0.5cm}
& \multicolumn{5}{l}{ $\lambda = 0.1 \ \longrightarrow$ } \\ 
$\downarrow \ \lambda = 0.5$
& \hspace{0.4cm}1\hspace{0.4cm} 
& \hspace{0.4cm}2\hspace{0.4cm}
& \hspace{0.4cm}3\hspace{0.4cm}
& \hspace{0.4cm}4\hspace{0.4cm} 
& \hspace{0.4cm}5\hspace{0.4cm}
& \hspace{0.4cm}6\hspace{0.4cm}
& \hspace{0.4cm}7\hspace{0.4cm} 
\\ \hline
\hspace{0.7cm} 1 & \underline{0.491} & \underline{0.420} & 0.001
& 0.000 & 0.001 & 0.000 & 0.002 \\
\hspace{0.7cm} 2 & \underline{0.402} & \underline{0.503} & 0.004
& 0.001 & 0.001 & 0.002 & 0.005 \\
\hspace{0.7cm} 3 & 0.002 & 0.001 & \underline{0.565} & 0.194
& 0.053 & 0.002 & 0.008 \\
\hspace{0.7cm} 4 & 0.000 & 0.000 & 0.121 & \underline{0.327}
& 0.177 & 0.222 & 0.028 \\
\hspace{0.7cm} 5 & 0.002 & 0.000 & 0.000 & 0.004 & 0.050 &
0.114 & \underline{0.562} \\
\hline\hline
\end{tabular}
\end{center}
\caption{ \label{table:N12.Norb26.Mtot130.D0.6.M3B0.5vs0.1.1LL}
Overlap matrix of $\lambda = 0.5$ (row) and 0.1 (column) for $N =
12$, $N_{orb} = 26$, $M = 130$, and $d = 0.6$. The underlined
numbers are the overlap between an edge state for $\lambda = 0.5$ and the
(likely) corresponding edge state for $\lambda = 0.1$.}
\end{table}

\subsection{The pure Coulomb case}
\label{app:evolution0.0}

We now move to the pure Coulomb case with $\lambda = 0$ and look for
the eigenstates with significant overlap with the edge states in the
$\lambda = 0.5$ system.  Surprisingly for $\Delta M = 1$, the 7th
lowest state has the largest overlap with the bosonic eigenstate in
the corresponding subspace for $\lambda = 0.5$
(Table~\ref{table:N12.Norb26.Mtot127.D0.6.M3B0.5vs0.0.1LL}).  The
overlap 0.403 is far from unity, but comparable to that between the
Pfaffian state and the ground state of the pure Coulomb system.  The
six lower eigenstates (with the notable exception of the state \#5),
which have negligible overlaps with the edge state, have nonetheless
significant overlaps with the lowest energy bulk excited states 2-5 for
$\lambda = 0.5$, indicating their bulk nature.  The complexity of the
Coulomb case is thus evident even for the $\Delta M = 1$ case.

\begin{table}
\begin{center}
\begin{tabular}{lcccccccc}
\hline\hline
\hspace{0.5cm}$\Delta M = 1$\hspace{0.5cm}
& \multicolumn{5}{l}{ $\lambda = 0.0 \ \longrightarrow$ } \\ 
$\downarrow \ \lambda = 0.5$
& \hspace{0.3cm}1\hspace{0.3cm} 
& \hspace{0.3cm}2\hspace{0.3cm}
& \hspace{0.3cm}3\hspace{0.3cm}
& \hspace{0.3cm}4\hspace{0.3cm} 
& \hspace{0.3cm}5\hspace{0.3cm}
& \hspace{0.3cm}6\hspace{0.3cm}
& \hspace{0.3cm}7\hspace{0.3cm} 
& \hspace{0.3cm}8\hspace{0.3cm}
\\ \hline
\hspace{0.7cm} 1 & 0.003 & 0.000 & 0.001 & 0.012 & 0.008 & 0.006 &
\underline{0.403} & 0.019 \\ \hspace{0.7cm} 2 & 0.309
& 0.281 & 0.007 & 0.047 & 0.004 & 0.012 & 0.006 & 0.007 \\ \hspace{0.7cm} 3 & 0.222 & 0.025 & 0.104 & 0.113 & 0.013 &
0.038 & 0.000 & 0.000 \\ \hspace{0.7cm} 4 & 0.007 &
0.036 & 0.103 & 0.034 & 0.012 & 0.170 & 0.000 & 0.001
\\ \hspace{0.7cm} 5 & 0.000 & 0.000 & 0.002 & 0.000 & 0.003 & 0.001 &
0.140 & 0.008 \\
\hline\hline
\end{tabular}
\end{center}
\caption{
\label{table:N12.Norb26.Mtot127.D0.6.M3B0.5vs0.0.1LL}
Overlap matrix of the two systems with
$\lambda = 0.5$ (row) and 0.0 (column) for $N = 12$,
$N_{orb} = 26$, $M = 127$, and $d = 0.6$.
The underlined element of 0.403 is the overlap of
the lowest state (edge state) for $\lambda = 0.5$ (mixed system) and
the seventh lowest state for $\lambda = 0.0$ (pure Coulomb system).
}
\end{table}

\begin{table}
\begin{center}
\begin{tabular}{lcccccc}
\hline\hline
\hspace{0.5cm}$\Delta M = 2$\hspace{0.5cm}
& \multicolumn{5}{l}{ $\lambda = 0.0 \ \longrightarrow$ } \\ 
$\downarrow \ \lambda = 0.5$
& \hspace{0.5cm}1\hspace{0.5cm} 
& \hspace{0.5cm}2\hspace{0.5cm}
& \hspace{0.5cm}3\hspace{0.5cm}
& \hspace{0.5cm}4\hspace{0.5cm} 
& \hspace{0.4cm}$\cdots$\hspace{0.4cm}
& \hspace{0.5cm}17\hspace{0.5cm} 
\\ \hline
\hspace{0.7cm}1 & 0.187 & \underline{0.432} & 0.070 & 0.000 & $\cdots$
& 0.000 \\
\hspace{0.7cm}2 & 0.003 & 0.001 & 0.000 & 0.000 & $\cdots$& \underline{0.168} \\
\hspace{0.7cm}3 & 0.000 & 0.000 & 0.000 & 0.000 & $\cdots$ &
0.000 \\
\hspace{0.7cm}4 & 0.288 & 0.206 & 0.003 & 0.019 & $\cdots$ &
0.007 \\
\hspace{0.7cm}5 & 0.005 & 0.015 & 0.006 & 0.187 & $\cdots$ & 0.000\\
\hline\hline
\end{tabular}
\end{center}
\caption{
\label{table:N12.Norb26.Mtot128.D0.6.M3B0.5vs0.0.1LL}
Overlap matrix of $\lambda = 0.5$ (row) and 0.0 (column) for $N = 12$,
$N_{orb} = 26$, $M = 128$, and $d = 0.6$. The
underlined numbers are the overlap between an edge state for $\lambda = 0.5$
and the (likely) corresponding edge state for the pure Coulomb case.}
\end{table}

\begin{table}
\begin{center}
\begin{tabular}{lcccccc}
\hline\hline
\hspace{0.5cm}$\Delta M = 3$\hspace{0.5cm}
& \multicolumn{5}{l}{ $\lambda = 0.0 \ \longrightarrow$ } \\ 
$\downarrow \ \lambda = 0.5$
& \hspace{0.5cm}1\hspace{0.5cm} 
& \hspace{0.5cm}2\hspace{0.5cm}
& \hspace{0.5cm}3\hspace{0.5cm}
& \hspace{0.5cm}4\hspace{0.5cm} 
& \hspace{0.4cm}$\cdots$\hspace{0.4cm}
& \hspace{0.5cm}17\hspace{0.5cm} 
\\ \hline
\hspace{0.7cm}1 & 0.220 & \underline{0.301} & 0.095
& 0.004 & $\cdots$ & 0.004 \\
\hspace{0.7cm}2 & 0.000 & 0.000 & 0.001
& 0.002 & $\cdots$ & \underline{0.361} \\
\hspace{0.7cm}3 & 0.012 & 0.014 & 0.003 & 0.001 & $\cdots$
& 0.001 \\
\hspace{0.7cm}4 & 0.000 & 0.000 & 0.000 & 0.000 & $\cdots$ & 0.001 \\
\hspace{0.7cm}5 & 0.000 & 0.000 & 0.000 & 0.000 & $\cdots$ & 0.000 \\
\hline\hline
\end{tabular}
\end{center}
\caption{
\label{table:N12.Norb26.Mtot129.D0.6.M3B0.5vs0.0.1LL}
Overlap matrix of $\lambda = 0.5$ (row) and 0.0 (column) for $N = 12$,
$N_{orb} = 26$, $M = 129$, and $d = 0.6$. The
underlined numbers are the overlap between an edge state for $\lambda = 0.5$
and the (likely) corresponding edge state for the pure Coulomb case.}
\end{table}

The attempt to find all edge states, even with the overlap matrix
calculation, is challenged by the following two difficulties.  First,
the edge states now have very large excitation energies, and thus a
lot more eigenstates are needed for the search.  Second, the overlaps
with eigenstates for $\lambda = 0.5$ fail to exhibit a clear
one-to-one correspondence.  In many cases, an edge states for $\lambda
= 0.5$ can have comparable overlaps with two eigenstates for $\lambda
= 0$, making the identification ambiguous. Despite the difficulties,
we can identify a number of edge states with some confidence.

We end the Appendix by making several observations. First, at this
relatively small system size, bulk excited states can have energies as
low as those of edge states.
In fact, the lowest energy eigenstates for $\Delta M =
1$-3 are bulk states, with small but finite overlaps ($\sim 0.2$) with
the corresponding edge states (see
Tables~\ref{table:N12.Norb26.Mtot127.D0.6.M3B0.5vs0.0.1LL},
\ref{table:N12.Norb26.Mtot128.D0.6.M3B0.5vs0.0.1LL} and
\ref{table:N12.Norb26.Mtot129.D0.6.M3B0.5vs0.0.1LL}).  This suggests
in the pure Coulomb case fermionic edge states mix with bulk states,
which is consistent with the fact that fermionic edge states
extrapolated from finite-$\lambda$ neutral velocities are expected at
energies in between the corresponding lowest two levels for $\Delta M
= 2$ and 3 (see Fig.~\ref{fig:edgemodes0.0}). 
A recent density-matrix renormalization group (DMRG)
calculation~\cite{feiguin07} estimate the excitation gap to be about
0.03 $e^2/\epsilon l_B$, thus we expect these bulk states will float
up in the thermodynamic limit.  Second, the low-lying fermionic edge
excitations do exist for $\lambda = 0$ at small excitation energies.
However, we cannot easily decompose these states into Majorana
fermionic levels with linear dispersion relation as we have done for
$\lambda = 0.5$ and 0.1.  The difficulty is due to mixing of the
fermionic edge excitations with bulk states.
Third,
there is significant redistribution in the weight of the lowest two edge
excitations for $\Delta M = 4$ as $\lambda$ decreases, as indicated by
the overlaps of the two states for $\lambda=0.1$ and 0.0 with those for
$\lambda=0.5$ (see
Tables~\ref{table:N12.Norb26.Mtot130.D0.6.M3B0.5vs0.1.1LL} and
\ref{table:N12.Norb26.Mtot130.D0.6.M3B0.5vs0.0.1LL}).

\begin{table}
\begin{center}
\begin{tabular}{lccccccc}
\hline\hline
\hspace{0.5cm}$\Delta M = 4$\hspace{0.5cm}
& \multicolumn{5}{l}{ $\lambda = 0.0 \ \longrightarrow$ } \\ 
$\downarrow \ \lambda = 0.5$
& \hspace{0.4cm}1\hspace{0.4cm} 
& \hspace{0.4cm}2\hspace{0.4cm}
& \hspace{0.4cm}3\hspace{0.4cm}
& \hspace{0.4cm}4\hspace{0.4cm} 
& \hspace{0.4cm}5\hspace{0.4cm}
& \hspace{0.4cm}6\hspace{0.4cm}
& \hspace{0.4cm}7\hspace{0.4cm} 
\\ \hline
\hspace{0.7cm} 1 & \underline{0.252} & \underline{0.250} & 0.004
& 0.003 & 0.008 & 0.005 & 0.088 \\
\hspace{0.7cm} 2 & \underline{0.284} & \underline{0.240} & 0.029
& 0.001 & 0.029 & 0.000 & 0.013 \\
\hspace{0.7cm} 3 & 0.002 & 0.000 & 0.029 & 0.005 & 0.045
& \underline{0.093} & 0.003 \\
\hspace{0.7cm} 4 & 0.001 & 0.000 & 0.019 & 0.000 & 0.023 & 0.005 &
0.000 \\
\hspace{0.7cm} 5 & 0.004 & 0.000 & 0.000 & 0.001 & 0.001 & 0.000 &
0.001 \\
\hline\hline
\end{tabular}
\end{center}
\caption{
\label{table:N12.Norb26.Mtot130.D0.6.M3B0.5vs0.0.1LL}
Overlap matrix of $\lambda = 0.5$ (row) and 0.0 (column) for $N = 12$,
$N_{orb} = 26$, $M = 130$, and $d = 0.6$. The
underlined numbers are the overlap between an edge state for $\lambda = 0.5$
and the (likely) corresponding edge state for the pure Coulomb case.}
\end{table}


\begin{thebibliography}{99}

\bibitem{wen95}
X.-G. Wen, Adv. Phys. {\bf 44}, 405 (1995).

\bibitem{fcharge} V. J. Goldman and B. Su,
Science {\bf 267}, 1010 (1995);
  L. Saminadayar, D. C. Glattli,
  Y. Jin and B. Etienne, Phys. Rev. Lett. {\bf 79}, 2526 (1997); R. de-Picciotto, M. Reznikov, M. Heiblum, V. Umansky, G. Bunin and D.
Mahalu, Nature (London) {\bf 389}, 162 (1997).

\bibitem{goldman}  F. E. Camino, W. Zhou, and V. J. Goldman,
Phys. Rev. B {\bf 72}, 075342 (2005).

\bibitem{chang03}
A.~M. Chang, Rev. Mod. Phys. {\bf 75}, 1449 (2003).

\bibitem{willett87}
R.~L. Willett, J.~P. Eisenstein, H.~L. Stormer, D.~C. Tsui, A.~C.
Gossard, and J.~H. English, Phys. Rev. Lett. {\bf 59}, 1776 (1987).

\bibitem{morf} R. H. Morf, Phys.
Rev. Lett. {\bf 80}, 1505 (1998).

\bibitem{rh} E.~H. Rezayi and
F.~D.~M. Haldane, Phys. Rev. Lett. {\bf 84}, 4685 (2000).

\bibitem{moore91}
G. Moore and N. Read, Nucl. Phys. B {\bf 360}, 362 (1991).

\bibitem{note} A working assumption here is that the lowest Landau
level has been fully occupied by both up- and down-spin electrons,
and this accounts for filling factor 2; the remaining electrons with
filling factor 1/2 occupy the first excited Landau level.

\bibitem{nayak96}
C. Nayak and F. Wilczek, Nucl. Phys. B {\bf 479}, 529 (1996).

\bibitem{milovanovic96}
M. Milovanovi\'c and N. Read, Phys. Rev. B {\bf 53}, 13~559 (1996).

\bibitem{kitaev03}
A. Kitaev, Ann. Phys. {\bf 303}, 2 (2003).

\bibitem{freedman02a}
M.~H. Freedman, A. Kitaev, and Z. Wang, Commun. Math. Phys. {\bf
227}, 587 (2002).

\bibitem{freedman02b}
M.~H. Freedman, M. Larsen, and Z. Wang, Commun. Math. Phys. {\bf 227}, 605
(2002).

\bibitem{dassarma05}
S. Das Sarma, M. Freedman, and C. Nayak, Phys. Rev. Lett. {\bf 94},
166802 (2005).

\bibitem{bonesteel05}
N.~E. Bonesteel, L. Hormozi, and G. Zikos, and S.~H. Simon, Phys.
Rev. Lett. {\bf 95}, 140503 (2005).

\bibitem{lee07}
S.-S. Lee, S. Ryu, C. Nayak, and M. P. A. Fisher, Phys. Rev. Lett.
{\bf 99}, 236807 (2007).

\bibitem{levin07}
M. Levin, B. I. Halperin, and B. Rosenow, Phys. Rev. Lett. {\bf 99},
236806 (2007).

\bibitem{miller07} J. B. Miller, I. P. Radu, D. M. Zumbuhl,
E. M. Levenson-Falk, M. A. Kastner, C. M. Marcus, L. N. Pfeiffer,
and K. W. West, Nature Physics {\bf 3}, 561 (2007).

\bibitem{fradkin98}
E. Fradkin, C. Nayak, A. Tsvelik, and F. Wilczek, Nucl. Phys. B {\bf
479}, 529 (1996).

\bibitem{stern06}
A. Stern and B.~I. Halperin, Phys. Rev. Lett. {\bf 96}, 016802
(2006).

\bibitem{bonderson06a}
P. Bonderson, A. Kitaev, and K. Shtengel, Phys. Rev. Lett. {\bf 96},
016803 (2006).

\bibitem{ardonne07} E. Ardonne and E.-A. Kim, arxiv:0705.2902.

\bibitem{overbosch07} B. J. Overbosch and X.-G. Wen, arxiv:0706.4339.

\bibitem{bishara07}
W. Bishara and C. Nayak, Phys. Rev. B {\bf 77}, 165302 (2008).

\bibitem{wan06}
X. Wan, K. Yang, and E.~H. Rezayi, Phys. Rev. Lett. {\bf 97}, 256804
(2006).

\bibitem{simonetal} S.~H. Simon, E.~H. Rezayi and N.~R. Cooper, Phys.
 Rev. B. {\bf 75}, 195306 (2007); {\it ibid.} {\bf 75}, 075318
 (2007); S.~H. Simon, E.~H. Rezayi, N.~R. Cooper, and I. Berdnikov,
 {\it ibid.} {\bf 75}, 075317 (2007).

\bibitem{haldane} B.~A. Bernevig and F.~D.~M. Haldane, arXiv:0711.3062.

\bibitem{wan03}
X. Wan, E.~H. Rezayi, and K. Yang, Phys. Rev. B {\bf 68}, 125307
(2003).

\bibitem{chamon94}
C. de C. Chamon and X.~G. Wen,
Phys. Rev. B {\bf 49}, 8227 (1994).

\bibitem{halperin83}
B.~I. Halperin, Helv. Phys. Acta. {\bf 56}, 75 (1983).

\bibitem{hu07}
Z.-X. Hu, X. Wan, and P. Schmitteckert, Phys. Rev. B 77, 075331 (2008).

\bibitem{footnote1} 
Two $+e/4$ quasiholes can form an alternative $+e/2$ quasihole, which
carries a neutral fermion (the $\psi$ field from the Ising conformal
field theory).~\cite{moore91} In the finite system, that state needs
an odd number of electrons to realize due to a global constraint (not
considered in this work). For even number of electrons, such a
$+e/2$ quaishole with a neutral fermion would require exciting an odd
number of fermion modes at the edge.

\bibitem{footnote}
On a sphere, the counterpart of this wave function is 
$$
{\rm Pf} \left ({u_iv_j + u_jv_i} \over {u_iv_j - u_jv_i} \right )
\prod_{i < j} (u_iv_j - u_jv_i)^2,
$$
where $(u,v)$ are the usual spinor coordinates.

\bibitem{toke}
C. T\H oke, N. Regnault, and J.~K. Jain,
Phys. Rev. Lett. {\bf 98},
036806 (2006); arXiv:0707.0586.

\bibitem{fendley} P. Fendley, M.~P.~A. Fisher, and C.  Nayak,
Phys. Rev. Lett. {\bf 97}, 036801 (2006); Phys. Rev. B {\bf 75},
045317 (2007).

\bibitem{palacios96}
J.~J. Palacios and A.~H. MacDonald, Phys. Rev. Lett. {\bf 76}, 118
(1996).

\bibitem{wan02}
X. Wan, K. Yang, and E.~H. Rezayi, Phys. Rev. Lett. {\bf 88}, 056802
(2002).

\bibitem{cazalilla05}
M.~A. Cazalilla, N. Barber\'an, and N.~R. Cooper, Phys. Rev. B {\bf
71}, 121303(R) (2005).

\bibitem{feiguin07}
A.~E. Feiguin, E.~H. Rezayi, C. Nayak, and S. Das Sarma,
arXiv:0706:4469v2.

\bibitem{yu07}
Y. Yu, J. Phys.:Condens. Matter {\bf 19}, 621 (2007).

\bibitem{lehur02}
K. Le Hur, Phys. Rev. B {\bf 65}, 233314 (2002); Phys. Rev. Lett. {\bf
95}, 076801 (2005); Phys. Rev. B {\bf 74}, 165104 (2006).

\bibitem{xia04}
J.~S. Xia, W. Pan, C.~L. Vicente, E.~D. Adams, N.~S. Sullivan, H.~L.
Stormer,
D.~C. Tsui, L.~N. Pfeiffer, K.~W. Baldwin, and K.~W. West, Phys. Rev. Lett.
{\bf 93}, 176809 (2004).

\bibitem{huber05}
M. Huber, M. Grayson, M. Rother, W. Biberacher, W. Wegscheider, and
G. Abstreiter, Phys. Rev. Lett. {\bf 94}, 016805 (2005).

\bibitem{dassarma07}
S. Das Sarma, M. Freedman, C. Nayak, S.~H. Simon, and A. Stern,
arXiv:0707.1889v1.

\bibitem{peterson08}
M.~R. Peterson and S. Das Sarma, arXiv:0801.4819.

\end{thebibliography}
\end{document}